\documentclass[reprint,aps,prc,twocolumn,showpacs,showkeys,10pt,longbibliography,
floatfix]{revtex4-2}

\usepackage{amsmath}
\usepackage{amsfonts}
\usepackage{graphicx}
\usepackage{float}
\usepackage{dcolumn}
\usepackage{multirow}
\usepackage{natbib}
\usepackage[colorlinks = true,linkcolor = blue,urlcolor  = blue,citecolor = blue,anchorcolor = blue]{hyperref}
\usepackage{txfonts}

\usepackage{microtype}

\begin{document}

\title{Crust-core interface and  bulk  neutron star properties}

\author{Ch. Margaritis}
\email{chmargar@auth.gr}

\author{P.S. Koliogiannis}
\email{pkoliogi@physics.auth.gr}

\author{A. Kanakis-Pegios}
\email{alkanaki@auth.gr}

\author{Ch.C. Moustakidis}
\email{moustaki@auth.gr}

\affiliation{Department of Theoretical Physics, Aristotle University of Thessaloniki, 54124 Thessaloniki, Greece}

\begin{abstract}
The nuclear symmetry energy plays an important role in the description of the properties of finite nuclei as well as neutron stars. Especially, for low values of baryon density, the accurate description of the crust-core interface strongly depends on the symmetry energy. Usually, the well known parabolic approximation is employed for the definition of the symmetry energy without avoiding some drawbacks. In the present paper, a class of nuclear models, suitable for the description of the inner and outer core of neutron stars, is applied in studying the effect of higher orders of the expansion of the energy on the location of the crust-core transition. The thermodynamical and dynamical methods are used for the determination of the transition density $n_{\rm t}$ and pressure $P_{\rm t}$. The corresponding energy density functional is applied for the study of some relevant properties of both nonrotating and slowly rotating neutron stars. We found that the larger the value of the slope parameter $L$, the slower the convergence of the expansion. In addition, a universal relation is presented between $n_{\rm t}$ and $L$, by employing the full expression and dynamical approach. The crustal moment of inertia is very sensitive to the location of the transition while the effects are moderated concerning the critical angular velocity of the $r$-mode instability and minimum mass configuration. The effect on the tidal deformability is less but not negligible. In any case, the use of the parabolic approximation leads to the overestimation of $n_{\rm t}$ and $P_{\rm t}$ and consequently, on inaccurate predictions. Moreover, in some cases, even  the matching process at the interface  may affect considerably the predictions, introducing errors of the same order with the one due to the employed method.    



\end{abstract}

\maketitle

\section{Introduction}
The equation of state (EoS) of neutron-rich nuclear matter is the main ingredient in the study  of the structure and properties of neutron stars~\cite{Shapiro-83,Glendenning-2000,Haensel-2007,Bertulani-2012,Lattimer-07}. Moreover, the observations of neutron stars provide useful constraints concerning the EoS both for low and high nuclear matter densities. In particular, of great interest are the properties related to the interface between the crust and core, including mainly the transition density and the corresponding transition pressure. These quantities are sensitive to the behavior of the EoS at low densities and play an important role in the predictions of some bulk neutron star properties. To be more specific, the inner crust comprises the outer region from the density at which neutrons drip out of nuclei to the inner edge, separating the solid crust from the homogeneous liquid core. At the inner edge a phase transition occurs from the high density homogeneous matter to the inhomogeneous one at lower densities. It was found that the transition density determines the structure of the inner part of the neutron star's crust and is also related to some finite nuclei properties including neutron skin, dipole polarizability, etc.~\cite{Paar-014,Centelles-09,Horowitz-2000}.

The determination of the transition density $n_{\rm t}$ itself is a very complicated problem because the inner crust may have an intricate structure. A well established approach is to find the density
at which the uniform liquid becomes unstable with respect to small-amplitude density fluctuations, indicating the formation of nuclear clusters. This approach includes mainly the dynamical method~\cite{Baym-71b,Pethick-95b,Oyamatsu-07,Ducoin-07,Xu-09,Lattimer-13,Feng-17,Tsaloukidis-2019,Gonzalez-2019,Ferreira-2020}, the thermodynamical one~\cite{Kubis-07a,Kubis-07b,Moustakidis-010,Moustakidis-012,Li-2020}, the random phase approximation~\cite{Horowitz-2000,Carriere-03}, and the Vlasov method~\cite{Pais-2010,Pais-2016}. Recently, in a notable work, Carreau \textit{et al.}~\cite{Carreau-2019} studied the crust-core transition within a unified meta-modeling of the nuclear EoS where the variational equations in the crust are solved within a compressible liquid-drop approach.

The motivation of the present paper is twofold. First, we intend to study more systematically the convergence of the baryon energy per particle expansion around the asymmetry parameter $I=(n_n-n_p)/(n_n + n_p)$, where $n_n$ and $n_p$ are the neutron and proton number densities, respectively (for a similar study see also   Refs~\cite{Zuo-2002,Routray-2016,Ducoin-2011,Cai-2012,Seif-2014,Boquera-2017,Tsukioka-2017,Vidana-2009,Sellahewa-2014}). It is  known that keeping only a quadratic term of $I$ (parabolic approximation), the accuracy of the expansion is sufficient. In particular,  some recent microscopic (\textit{ab initio}) calculations reinforce  the accuracy of the  parabolic approximation even for low values of the baryon density. The authors in Ref.~\cite{Zuo-2002} found that despite the strongly density-dependent repulsive effect, the use of the three body forces in the framework of the Brueckner-Bethe-Goldstone theory does not violate the parabolic law, which is already fulfilled with the corresponding two body forces. Finally, their conclusion for a desirable result is based on (a) the agreement with previous similar studies, which provided a strong support for using the parabolic approximation to describe isospin effects, and (b) the strict theoretical constraint that imposes on the phenomenological nuclear models when extended to asymmetric nuclear matter (i.e. Skyrme forces)~\cite{Zuo-2002}. In addition, the authors in Ref.~\cite{Vidana-2009} performed a systematic analysis of the density dependence of nuclear symmetry energy within the microscopic Brueckner-Hartree-Fock approach using the realistic Argonne V18 potential plus phenomenological three body forces of Urbana type. They found that the results of their microscopic model are well reproduced by the parabolic approximation, at least in the range of the asymmetry that is taken into account.

In recent studies, the effects of additional terms have been considered, usually up to the order of ${\cal O}(I^6)$~\cite{Boquera-2017}. In the present paper we extend the study up to the order of ${\cal O}(I^{10})$, conjecturing that it is sufficient to gain the main conclusion about the speed of convergence of the expansion. In particular, we employ the energies per baryon, originated by various nuclear models, mainly focusing on the role played by the slope of the symmetry energy $L$ at the saturation density and its effect on the growth of the convergence speed. Moreover, we employed both the dynamical and thermodynamical method to calculate the transition density and pressure, corresponding to the crust-core interface for each expansion term. We mainly focus on the parabolic approximation (PA) and the full expression (FE) (which includes all terms).

Secondly, we apply our findings in order to examine the effect of the two cases on some neutron star properties that we expect to be sensitive to the values of $n_{\rm t}$ and $P_{\rm t}$. To be more specific, we concentrate on the crustal moment of inertia both of nonrotating and of slowly rotating neutron stars, the critical frequency of the $r$-mode instability, the tidal deformability, and the minimum mass configuration. Finally, we analyze and discuss the extent of the effects of the crust-core interface, related to the specific choice of the applied method and matching process, on each of the latter properties.

The paper is organized as follows: In Sec.~\ref{sec:sym_energy} we present the energy expansion, the symmetry energy, and the corresponding slope parameter. In Sec.~\ref{sec:ther_dyn_method} we briefly summarize the dynamical and the thermodynamical method for the determination of the transition density and pressure, while in Sec.~\ref{sec:applications} we present the specific neutron star properties related to the crust-core interface. In Sec.~\ref{sec:results} we present the results of the present paper and discuss their implications. Finally, Sec.~\ref{sec:remarks} includes the concluding remarks, and Appendix provides the thermodynamically consistent matching process.

\section{Symmetry energy} \label{sec:sym_energy}
The energy per baryon $E_b(n,I)$ of asymmetric nuclear matter can be expanded around the asymmetry parameter $I$  as~\cite{Moustakidis-012,Li-2014}
\begin{eqnarray}
	E_b(n,I) &=& E_b(n,I=0) + \sum_{k=1}E_{{\rm sym},2k}(n)I^{2k}.
	\label{esym-exp-1}
\end{eqnarray}
The asymmetry parameter can be written as $I= 1-2x$, where $x$ is the proton
fraction $n_p/n$. Moreover, in Eq.~\eqref{esym-exp-1}, $E_b(n,I=0)$ denotes the energy per baryon of the symmetric nuclear matter. The coefficients of the expansion in Eq.~\eqref{esym-exp-1} are given by the expression
\begin{equation}
	\left.E_{{\rm sym},2k}(n)=\frac{1}{(2k)!}\frac{\partial^{2k}E_b(n,I)}{\partial I^{2k}}\right\vert_{I=0}.
	\label{Esy-coef}
\end{equation}
Since the strong interaction must be symmetric under the exchange of neutrons with protons, only even powers of $I$ appear in Eq.~\eqref{esym-exp-1}. The nuclear symmetry energy is, by definition, the coefficient of the quadratic term $E_{\rm sym,2}(n)$. The slope parameter $L$ is an indicator of the stiffness of the EoS and is defined as
\begin{equation}
	L=3n_s\left.\frac{dE_{\rm sym,2}(n)}{d n}\right\vert_{n=n_s},
	\label{L-def}
\end{equation}
where $n_s$ is the nuclear saturation density.

We can define also the slope parameter $L_{2k}$, which corresponds to higher order terms, using Eq.~\eqref{Esy-coef} (see also Ref.~\cite{Boquera-2017}), according to the rule
\begin{equation}
	L_{2k}=3n_s\left.\frac{dE_{{\rm sym},2k}(n)}{d n}\right\vert_{n=n_s}.
	\label{L-sym}
\end{equation}  
Keeping in Eq.~\eqref{esym-exp-1} terms only up to a quadratic one, and defining the symmetry energy as the difference between the energy per baryon in pure neutron matter and symmetric nuclear matter, we are lead to the parabolic approximation of the symmetry energy according to the law
\begin{equation}
	E^{\rm PA}_{\rm sym}(n)=E_b(n,I=1)-E_b(n,I=0).
	\label{E-pa}
\end{equation}
The slope parameter $L_{\rm PA}$, that corresponds to the parabolic approximation, is found by Eq.~\eqref{L-def} by replacing $E_{{\rm sym},2}(n)$ with $E^{\rm PA}_{\rm sym}(n)$. It is worth mentioning here that in general, the definitions of $E_{{\rm sym},2}(n)$ and $E^{\rm PA}_{\rm sym}(n)$ do not coincide. This is the case only when the energy per baryon includes terms up to a quadratic one of the asymmetry parameter $I$.   

\section{Dynamical and thermodynamical method} \label{sec:ther_dyn_method}
The crust-core interface is related to the phase transition between nuclei and uniform nuclear matter. The latter one is nearly pure neutron matter, as the proton fraction is just a few percent, determined by the condition of $\beta$ equilibrium. The study of the instability of $\beta$-stable nuclear matter is based on the variation of the total energy density in the framework of the Thomas-Fermi approximations~\cite{Baym-71b,Pethick-95b}. In the dynamical method, compared to the thermodynamical one, effects due to inhomogeneity of the density and the Coulomb interaction have also been included. The onset of instability will occur if the total energy, in the presence of the density inhomogeneity, is lower than the energy of the uniform phase. The key expression for the description of the instability reads as~\cite{Baym-71b,Pethick-95b,Tsaloukidis-2019}
\begin{equation}
	U_{\rm dyn}(n)=U_0(n)+4\sqrt{\pi \alpha \hbar c\xi}-4 \alpha \xi  \left(9\pi x^2 n^2 \right)^{1/3}, 
	\label{VQ-min}
\end{equation}
where $\alpha=e^2/\hbar c$,
\begin{equation}
	U_0(n)=\frac{\partial \mu_p}{\partial n_p}-\frac{(\partial \mu_p/\partial n_n)^2}{\partial \mu_n/\partial n_n}, 
	\label{V0}
\end{equation}
and 
\begin{equation}
	\xi=2D_{nn}(1+4\zeta +\zeta^2),  \quad \zeta=-\frac{\partial \mu_p/\partial n_n}{\partial \mu_n/\partial n_n}.
	\label{xi-1}
\end{equation}
The  chemical potentials  $\mu_n$ and $\mu_p$ are defined as
\begin{equation}
	\mu_n=\left(\frac{\partial E_b}{\partial n_n}  \right)_{n_p}, \quad \mu_p=\left(\frac{\partial E_b}{\partial n_p}  \right)_{n_n}.
	\label{Def-chem-pot}
\end{equation}
The transition density $n_{\rm t}$ is determined from the condition $U_{\rm dyn}(n_{\rm t})=0$.

The key expression of the thermodynamical method is the following (for more details see Refs.~\cite{Kubis-07a,Kubis-07b,Moustakidis-010,Moustakidis-012}),
\begin{eqnarray}
	C_{\rm therm}(n)&=&2n\frac{\partial E(n,x)}{\partial n}+n^2\frac{\partial^2 E(n,x)}{\partial n^2}\nonumber \\
	&-&\left(\frac{\partial^2 E(n,x)}{\partial n \partial x}n  \right)^2\left(\frac{\partial^2 E(n,x)}{\partial x^2}  \right)^{-1},
	\label{therm-cond-1}
\end{eqnarray}
and the transition density $n_{\rm t}$ is now determined from the condition $C_{\rm therm}(n_{\rm t})=0$.

In both cases, the proton fraction $x$ is determined as a function of the baryon density $n$ from the condition of $\beta$ equilibrium. In particular, in $\beta$-stable nuclear matter, the chemical equilibrium condition takes the form
\begin{equation}
	\mu_n=\mu_p+\mu_e.
	\label{beta-1}
\end{equation}
It is easy to show that after some algebra we get~\cite{Moustakidis-012,Tsaloukidis-2019} 
\begin{equation}
	\mu_n-\mu_p=\left(-\frac{\partial E_b}{\partial x}\right)_n.
	\label{beta-2}
\end{equation} 
Now, since the electron chemical potential $\mu_e$ is given by
\begin{equation}
	\mu_e=\hbar c (3\pi^2 n_e)^{1/3},
	\label{mu-e-1}
\end{equation}
we finally find
\begin{equation}
	\left(\frac{\partial E_b}{\partial x}\right)_n=-\hbar c (3\pi^2 x n)^{1/3}.
	\label{beta-3}
\end{equation}
Eq.~\eqref{beta-3} is solved numerically and an expression for $x$ as a function of $n$ is found.

Another important quantity is the transition pressure $P_{\rm t}$. Both baryons and leptons contribute to the total pressure, meaning that the transition pressure is given by the following relation~\cite{Moustakidis-010,Moustakidis-012}:
\begin{equation}
	P_{\rm t}^{\rm FE}(n_{\rm t},x_{\rm t})=n_{\rm t}^2\left.\frac{\partial E_b}{\partial
	n}\right\vert_{n=n_{\rm t}}+\frac{\hbar c}{12 \pi^2}\left(3\pi^2
	x_{\rm t}n_{\rm t}\right)^{4/3}, 
	\label{Pr-tra}
\end{equation}
where $x_{\rm t}$ is the proton fraction related to the transition density.

\section{Application on neutron star properties} \label{sec:applications}
The EoS of nuclear matter is the key ingredient to study the bulk properties of neutron stars. However, there are some specific ones which depend directly on the location of the crust-core interface. This class of properties includes the fraction of the moment of inertia of the crust with respect to the total one, the critical angular frequency which defines the $r$-mode instability, the tidal polarizability, and the minimum mass configuration. To be more specific, the crustal moment of inertia $I_{\rm crust}/I$  and the critical angular velocity $\Omega_{c}$, by definition, depend on the crust-core interface. On the other hand, tidal deformability $\lambda$ of a low mass neutron star is sensitive to the specific details of the crust~\cite{Perot-2020,Piekarewicz-2019,Kalaitzis-2019,Gittins-2020}, while  minimum mass configuration mainly depends on the contribution of the crust both to $M_{\rm min}$ and to $R_{\rm min}$. In the latter, since the central density is close to the transition density, it is expected that even a slight shift to $n_{\rm t}$ (and $P_{\rm t}$) will modify the relevant predictions.

\subsection{Crustal fraction of the moment of inertia}
The crustal moment of inertia plays an important role in the evolution of neutron stars and the exhibition of some specific phenomena. It is particularly interesting since it can be inferred from observations of pulsar glitches, the occasional disruptions of the otherwise extremely regular pulsations of magnetized, rotating neutron stars. In the case of a nonrotating (or slowly rotating) neutron star a few elaborated approximations have been performed leading to analytical predictions of $I_{\rm crust}$. These approximations depend directly on the transition pressure $P_{\rm t}$ (or/and  the transition density $n_{\rm t}$). The most used one was provided in Ref.~\cite{Link-99} and is written as   
\begin{eqnarray}
	\frac{I_{\rm crust}}{I}&\simeq& \frac{28 \pi P_{\rm t}R^3}{3Mc^2}\frac{(1-1.67\beta-0.6\beta^2)}{\beta}\nonumber\\
	&\times&\left(1+\frac{2P_{\rm t}}{n_{\rm t}mc^2}\frac{(1+5\beta-14\beta^2)}{\beta^2}  \right)^{-1},
	\label{crust-1}
\end{eqnarray}
where $\beta=GM/Rc^2$ is the compactness parameter.

We also apply for comparison a second approximation~\cite{Fattoyev-2010} given by
\begin{eqnarray}
	I_{\rm crust}&\simeq&\frac{16 \pi}{3}\frac{R_{\rm core}^6}{2\beta c^{2} R}P_{\rm t}\left(1-\frac{0.21}{1-2\beta}2\beta  \right)\nonumber \\
	&\times& \left[1+\frac{48}{5}\left(\frac{R_{\rm core}}{2\beta R}-1\right)\left(\frac{P_{\rm t}}{\cal {E}_{\rm t}}  \right)+\cdots\right].
	\label{crust-2}
\end{eqnarray}

In general, the moment of inertia of rotating neutron stars exhibits dependence on the spin frequency~\cite{Koliogiannis-2020} and in this case, the calculation of the crustal moment of inertia demands special treatment. However, it is worth noticing that approximations~\eqref{crust-1} and~\eqref{crust-2} are sufficiently accurate also for a slowly rotating neutron star, that is rotating with angular velocity $\Omega \ll \Omega_{\rm k}$, where $\Omega_{\rm k}$ is the Kepler angular velocity~\cite{Fattoyev-2010}.

\subsection{Critical angular velocity for $r$-modes}
The $r$ modes are oscillations of rotating stars the restoring force of which is the Coriolis force (see Refs.~\cite{Andersson-2001,Friedman-99,Andersson-99} and references therein). The gravitational
radiation-driven instability of these models has been proposed as an explanation for the observed relatively low spin frequencies of young neutron stars and of accreting ones in low-mass x-ray binaries (LMXBs). This instability can only occur when the gravitational-radiation driving time scale of the $r$-mode is shorter than the time scales of the various dissipation mechanisms that may occur in the interior of the neutron star. The instability condition reads as~\cite{Andersson-2001,Friedman-99,Andersson-99,Moustakidis-2015}   
\begin{equation}
	\frac{1}{\tau_{\rm GW}}+\frac{1}{\tau_{ee}}+\frac{1}{\tau_{nn}}=0,
	\label{r-mode-1}
\end{equation}
where $\tau_{ee}$ and $\tau_{nn}$ are the time scales of the various dissipation mechanisms (due to electron-electron and neutron-neutron scattering respectively) which are considered in the present paper. The condition~\eqref{r-mode-1} leads to the critical angular velocity $\Omega_c$ which is given by (for a detailed analysis see Refs.~\cite{Tsaloukidis-2019,Moustakidis-2015,Zhou-2021})
\begin{equation}
	\frac{\Omega_c}{\Omega_0}=\left(-\frac{\tilde{\tau}_{\rm GW}(\tilde{\tau}_{ee}+\tilde{\tau}_{ee}) }{\tilde{\tau}_{ee} \tilde{\tau}_{nn}}  \right)^{2/11}\left(\frac{10^8 {\rm K}}{T}   \right)^{2/11}, 
	\label{r-mode-2}
\end{equation}
where
$\Omega_0=\sqrt{3  G M/4R^3}$, $T$ is the temperature, and also 
\[\tau_{\rm GW}=\tilde{\tau}_{\rm GW}\left(\frac{\Omega_0}{\Omega} \right)^6,\quad \tau_{ii}=\tilde{\tau}_{ii}\left(\frac{\Omega_0}{\Omega} \right)^{1/2}\left(\frac{10^8 {\rm K}}{T}   \right),   \]
with $ii=ee,nn$. In the present paper we will concentrate on the case where the main damping mechanism is due to the viscous dissipation at the boundary layer of the perfectly rigid crust and fluid core. In this case, the corresponding critical angular velocity takes the form~\cite{Tsaloukidis-2019,Moustakidis-2015}
\begin{eqnarray}
	\Omega_{c}&=&1.93795 \times 10^5 \left(\frac {R_{\rm core}}{{\rm km}}  \right)^{12/11}\left(\frac{{\cal E}_{\rm t}}{{\rm MeV~fm^{-3}}}  \right)^{3/11}
	\nonumber \\
	&\times&\left[1+0.25865\left(\frac{{\cal E}_{\rm t}}{{\rm MeV~fm^{-3}}}  \right)^{1/8}  \right]^{2/11}
	\nonumber \\
	&\times& 
	I(R_{\rm core})^{-4/11} \left(\frac{10^8 {\rm K}}{T}   \right)^{2/11}.
	\label{r-mode-3}
\end{eqnarray}
From Eq.~\eqref{r-mode-3} is obvious the direct dependence of $\Omega_{c}$ on the crust-core interface via ${\cal E}_{\rm t}$, as well as the indirect one via the values of the core radius $R_{\rm core}$ and the integral $I(R_{\rm core})$ 
where
\begin{equation}
	\hspace{-0.5cm}
	I(R_{\rm core})=\int_{0}^{R_{\rm core}}
	\left( \frac{{\cal E}(r)}{{\rm MeV \ fm^{-3}}} \right) \left(\frac{r}{{\rm km}}  \right)^6 d\left(\frac{r}{{\rm km}}  \right),
	\label{Ic-1}
\end{equation}
with ${\cal E}(r)$ being the energy density of neutron star matter at distance $r$ from the center.

\subsection{Tidal deformability }

\begin{table*}
	\caption{Transition density (in units of ${\rm fm}^{-3}$) and pressure (in units of ${\rm MeV~fm}^{-3}$) calculated using the full expression for each nuclear model ($n_{\rm t}^{\rm FE},P_{\rm t}^{\rm FE}$), the parabolic approximation ($n_{\rm t}^{\rm PA},P_{\rm t}^{\rm PA}$), and approximations of each EoS based on Eq.~\eqref{esym-exp-1} up to order $2k$ ($n_{\rm t,2k},P_{\rm t,2k}$). All calculations are performed in the framework of the thermodynamical method.}
	\begin{ruledtabular}
		\begin{tabular}{l|cccccccccccccc}
			Nuclear model & $n_{\rm t}^{\rm FE}$ & $P_{\rm t}^{\rm FE}$ & $n_{\rm t}^{\rm PA}$ & $P_{\rm t}^{\rm PA}$ & $n_{\rm t,2}$ & $P_{\rm t,2}$ & $n_{\rm t,4}$ & $P_{\rm t,4}$ & $n_{\rm t,6}$ & $P_{\rm t,6}$ & $n_{\rm t,8}$ & $P_{\rm t,8}$ & $n_{\rm t,10}$ & $P_{\rm t,10}$ \\
			\hline
			MDI(65) & 0.078 & 0.317 & 0.097 & 0.594 & 0.094 & 0.641 & 0.092 & 0.519 & 0.090 & 0.477 & 0.088 & 0.449 & 0.086 & 0.428 \\
			MDI(72.5) & 0.073 & 0.315 & 0.094 & 0.728 & 0.094 & 0.697 & 0.090 & 0.618 & 0.087 & 0.561 & 0.085 & 0.519 & 0.083 & 0.488\\
			MDI(80) & 0.068 & 0.286 & 0.094 & 0.836 & 0.095 & 0.783 & 0.090 & 0.684 & 0.086 & 0.608 & 0.084 & 0.553 & 0.082 & 0.513 \\
			MDI(95-30)& 0.058 & 0.146 & 0.099 & 1.078 & 0.099 & 1.071 & 0.093 & 0.845 & 0.087 & 0.697 & 0.083 & 0.601 & 0.081 & 0.532 \\
			MDI(95-32)& 0.064 & 0.277 & 0.097 & 1.054 & 0.097 & 1.026 & 0.091 & 0.857 & 0.087 & 0.742 & 0.084 & 0.662 & 0.081 & 0.604\\
			MDI(100) & 0.053 & 0.088 & 0.102 & 1.202 & 0.103 & 1.247 & 0.096 & 0.967 & 0.089 & 0.765 & 0.085 & 0.641 & 0.081 & 0.554\\
			MDI(110) & 0.047 & 0.047 & 0.106 & 1.484 & 0.109 & 1.723 & 0.108 & 1.548 & 0.094 & 0.993 & 0.088 & 0.799 & 0.083 & 0.658\\
			HLPS(49.4)& 0.089 & 0.551 & 0.097 & 0.694 & 0.098 & 0.675 & 0.095 & 0.651 & 0.094 & 0.630 & 0.093 & 0.614 & 0.092 & 0.602\\
			HLPS(29.5)& 0.098 & 0.455 & 0.104 & 0.495 & 0.105 & 0.513 & 0.103 & 0.509 & 0.102 & 0.500 & 0.101 & 0.492 & 0.100 & 0.485\\
			SkI4(60.4) & 0.081 & 0.337 & 0.091 & 0.496 & 0.091 & 0.481 & 0.089 & 0.453 & 0.087 & 0.432 & 0.086 & 0.416 & 0.085 & 0.404\\
			Ska(76.1) & 0.079 & 0.528 & 0.093 & 0.866 & 0.094 & 0.809 & 0.090 & 0.762 & 0.088 & 0.713 & 0.086 & 0.677 & 0.085 & 0.649\\
			Sly4(46) & 0.088 & 0.463 & 0.094 & 0.546 & 0.094 & 0.528 & 0.093 & 0.517 & 0.092 & 0.506 & 0.091 & 0.497 & 0.090 & 0.491 \\ 
		\end{tabular}
	\end{ruledtabular}
	\label{tab:1}
\end{table*}

\begin{table*}
	\caption{Transition density (in units of ${\rm fm}^{-3}$) and pressure (in units of ${\rm MeV~fm}^{-3}$) calculated using the full expression for each nuclear model ($n_{\rm t}^{\rm FE},P_{\rm t}^{\rm FE}$), the parabolic approximation ($n_{\rm t}^{\rm PA},P_{\rm t}^{\rm PA}$), and approximations of each EoS based on Eq.~\eqref{esym-exp-1} up to order $2k$ ($n_{\rm t,2k},P_{\rm t,2k}$). All calculations are performed in the framework of the dynamical method.}
	\begin{ruledtabular}
		\begin{tabular}{l|cccccccccccccc}
			Nuclear model & $n_{\rm t}^{\rm FE}$ & $P_{\rm t}^{\rm FE}$ & $n_{\rm t}^{\rm PA}$ & $P_{\rm t}^{\rm PA}$ & $n_{\rm t,2}$ & $P_{\rm t,2}$ & $n_{\rm t,4}$ & $P_{\rm t,4}$ & $n_{\rm t,6}$ & $P_{\rm t,6}$ & $n_{\rm t,8}$ & $P_{\rm t,8}$ & $n_{\rm t,10}$ & $P_{\rm t,10}$ \\
			\hline
			MDI(65) & 0.070 & 0.232 & 0.086 & 0.425 & 0.084 & 0.488 & 0.082 & 0.365 & 0.080 & 0.339 & 0.077 & 0.322 & 0.077 & 0.309\\
			MDI(72.5) & 0.064 & 0.212 & 0.082 & 0.483 & 0.083 & 0.458 & 0.079 & 0.413 & 0.077 & 0.380 & 0.075 & 0.354 & 0.074 & 0.334\\
			MDI(80) & 0.060 & 0.181 & 0.082 & 0.529 & 0.082 & 0.492 & 0.078 & 0.441 & 0.076 & 0.396 & 0.074 & 0.363 & 0.072 & 0.337\\
			MDI(95-30)& 0.050 & 0.075 & 0.084 & 0.615 & 0.084 & 0.602 & 0.079 & 0.479 & 0.075 & 0.400 & 0.072 & 0.347 & 0.069 & 0.309\\
			MDI(95-32)& 0.055 & 0.157 & 0.082 & 0.641 & 0.083 & 0.618 & 0.078 & 0.524 & 0.075 & 0.457 & 0.072 & 0.409 & 0.070 & 0.374\\
			MDI(100) & 0.047 & 0.039 & 0.085 & 0.656 & 0.086 & 0.670 & 0.079 & 0.508 & 0.075 & 0.408 & 0.072 & 0.343 & 0.069 & 0.298\\
			MDI(110) & 0.042 & 0.016 & 0.088 & 0.804 & 0.090 & 0.919 & 0.085 & 0.699 & 0.077 & 0.488 & 0.072 & 0.392 & 0.069 & 0.325\\
			HLPS(49.4)& 0.079 & 0.415 & 0.087 & 0.525 & 0.087 & 0.509 & 0.085 & 0.493 & 0.084 & 0.478 & 0.083 & 0.466 & 0.082 & 0.457\\
			HLPS(29.5)& 0.091 & 0.339 & 0.093 & 0.366 & 0.094 & 0.403 & 0.093 & 0.376 & 0.092 & 0.376 & 0.091 & 0.371 & 0.090 & 0.399\\
			SkI4(60.4) & 0.073 & 0.248 & 0.082 & 0.356 & 0.082 & 0.343 & 0.080 & 0.327 & 0.079 & 0.314 & 0.078 & 0.304 & 0.077 & 0.296\\
			Ska(76.1) & 0.069 & 0.377 & 0.082 & 0.622 & 0.083 & 0.580 & 0.080 & 0.553 & 0.078 & 0.520 & 0.077 & 0.494 & 0.075 & 0.474\\
			Sly4(46) & 0.080 & 0.365 & 0.085 & 0.427 & 0.085 & 0.411 & 0.083 & 0.405 & 0.083 & 0.398 & 0.082 & 0.392 & 0.082 & 0.387\\ 
		\end{tabular}
	\end{ruledtabular}
	\label{tab:2}
\end{table*}

Gravitational waves from the final stages of inspiraling binary neutron stars are one of the most important sources for ground-based gravitational wave detectors~\cite{Abbott-2018,Abbott-2019,Abbott-2020}. Flanagan and Hinderer~\cite{Flanagan-2008} have pointed out that tidal effects are also potentially measurable during the early part of the evolution when the waveform is relatively clean. The tidal fields induce quadrupole moments on the neutron stars. The response of the neutron star is described by the dimensionless so-called Love number $k_2$ which depends on the structure of the neutron star (both core and crust). The Love number is linearly related to the tidal deformability $\lambda$ according to 
$\lambda=2R^5k_2/3G$. Now, $k_2$ is given by~\cite{Postnikov-2010}
\begin{widetext}
\begin{eqnarray}
	k_2&=&\frac{8\beta^5}{5}\left(1-2\beta\right)^2\left[2-y_R+(y_R-1)2\beta \right]
	\left\{\frac{}{} 2\beta \left(6  -3y_R+3\beta (5y_R-8)\right) \right. \nonumber \\
	&+& 4\beta^3 \left.  \left(13-11y_R+\beta(3y_R-2)+2\beta^2(1+y_R)\right)\frac{}{}
	+ 3\left(1-2\beta \right)^2\left[2-y_R+2\beta(y_R-1)\right] {\rm ln}\left(1-2\beta\right)\right\}^{-1},
	\label{k2-def}
\end{eqnarray}
\end{widetext}
where the quantity $y_R\equiv y(R)$ is determined by solving the relevant differential equation of $y(r)$ simultaneously with the Tolman-Oppenheimer-Volkoff equations~\cite{Postnikov-2010}.

The chirp mass ${\cal M}_{c}$, which is one of the binary parameters that is well constrained by the gravitational wave detectors, is defined as~\cite{Abbott-2018,Abbott-2019}
\begin{equation}
    {\cal M}_c=\frac{(m_1 m_2)^{3/5}}{(m_1+m_2)^{1/5}}=m_1\frac{q^{3/5}}{(1+q)^{1/5}},
    \label{Chirp-1}
\end{equation}
where $m_1$ is the mass of the heavier component star and $m_2$ is that of the lighter one. Hence, the binary mass ratio $q=m_2/m_1$ is within $0< q \leq 1$. Moreover, the information about the tidal effects in a binary system which is transferred by the gravitational waves is characterized by the effective tidal deformability~\cite{Abbott-2018,Abbott-2019}
\begin{equation}
 \tilde{\Lambda}=\frac{16}{13}\frac{(12q+1)\Lambda_1+(12+q)q^4\Lambda_2}{(1+q)^5},
 \label{Lambda-tilde}
\end{equation}
where the dimensionless tidal deformability $\Lambda_{ i}$ is defined as
\begin{equation}
    \Lambda_{ i}=\frac{2}{3}k_2\left(\frac{R_i c^2}{M_iG}  \right)^5, \quad { i}=1,2.   
    \label{Lambda-dim}
\end{equation}
It is noted that the measure of the amplitude of the radiated gravitational waves provides information for the tidal deformability and consequently useful constraints on bulk neutron star properties, including mainly the radius. We expect that due to the strong dependence of $\tilde{\Lambda}$ on $R_{i}$ and also on the Love number $k_2$, the effects of the crust-core interface may affect its value. In any case, this possibility is of interest and also worth consideration.

\subsection{Minimum mass configuration}
The minimum neutron star mass, apart for the maximum one, is also of great interest in astrophysics~\cite{Haensel-2002,Colpi-89}. Its knowledge is related to the case of a neutron star in
a close binary system with a more compact partner (neutron star or black hole). In particular, the lower mass neutron star transfers mass to the more massive object, a process which ultimately leads to approaching its minimum value. Finally, crossing this value, the neutron star reaches a non equilibrium configuration. The minimum mass is a universal feature, independent of the details of the EoS and well constrained to the value $M_{\rm min}\simeq 0.1 \ M_{\odot}$. This is because the corresponding central densities are close to the values  of the transition densities $n_{\rm t}$. Now, since the equation of the crust is well known, all theoretical predictions for $M_{\rm min}$ converge. However, the corresponding radius $R_{\rm min}$ is very sensitive to the details of the EoS. We expect that the location of the crust-core transition will affect appreciably the values of $R_{\rm min}$. In the present paper we investigate to what extent $R_{\rm min}$ is affected by the values of $n_{\rm t}$ (and $P_{\rm t}$).
 
\section{Results and discussion} \label{sec:results}

\begin{figure*}
	\centering
	\includegraphics[width=7.6cm, height=5.6cm]{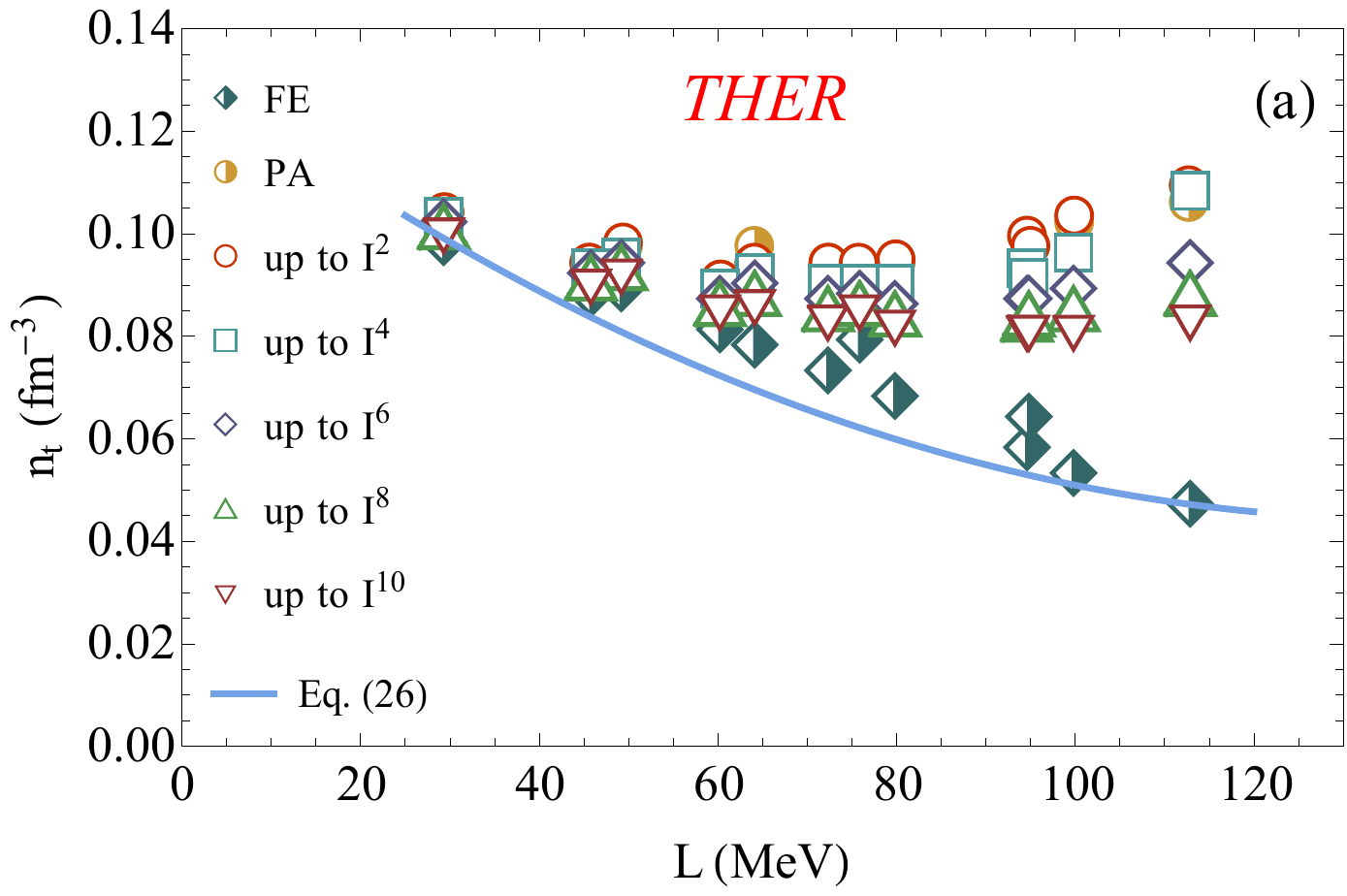}
	\hspace{1cm}
	\includegraphics[width=7.6cm, height=5.4cm]{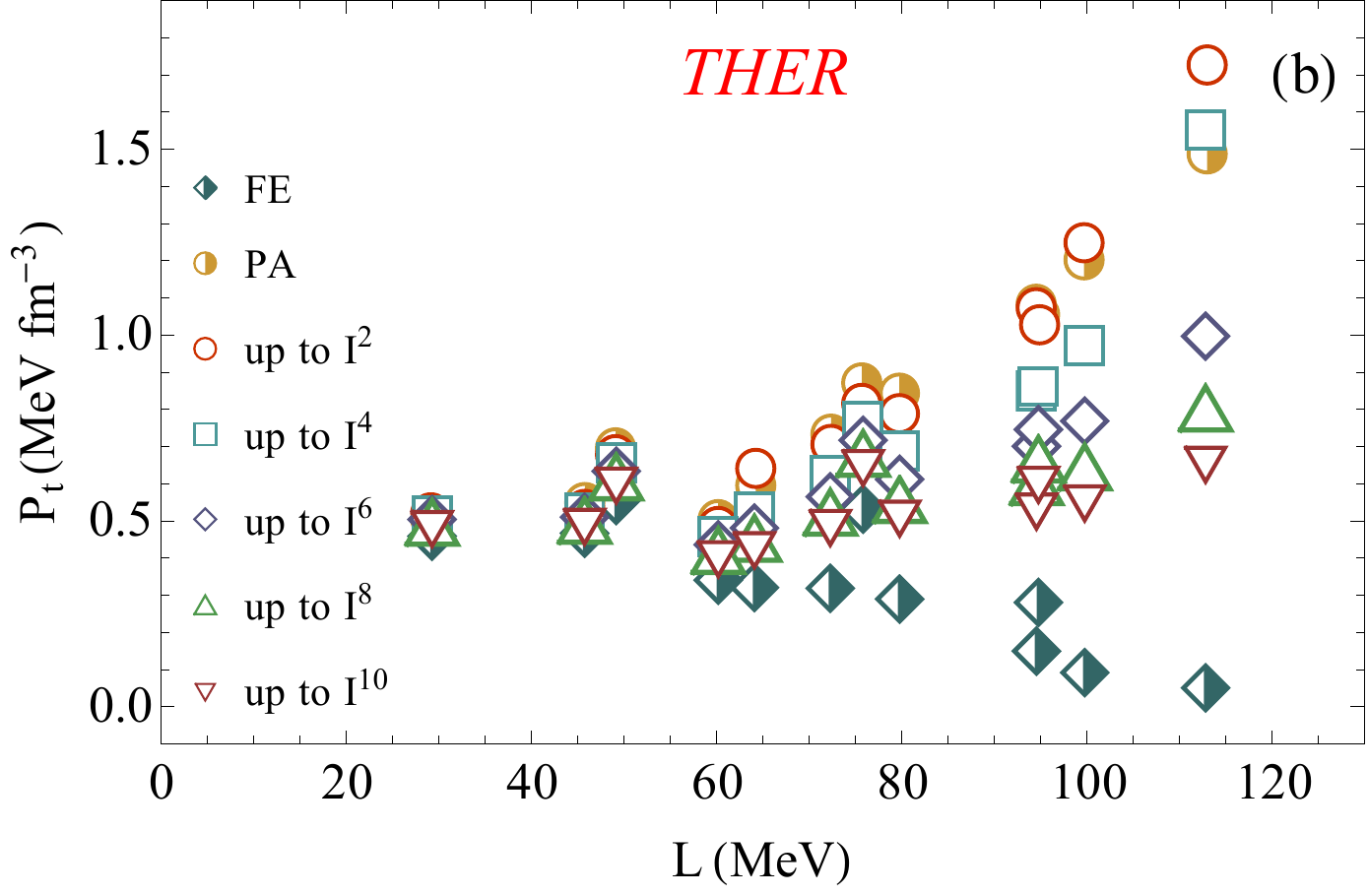}
	
	\includegraphics[width=7.6cm, height=5.4cm]{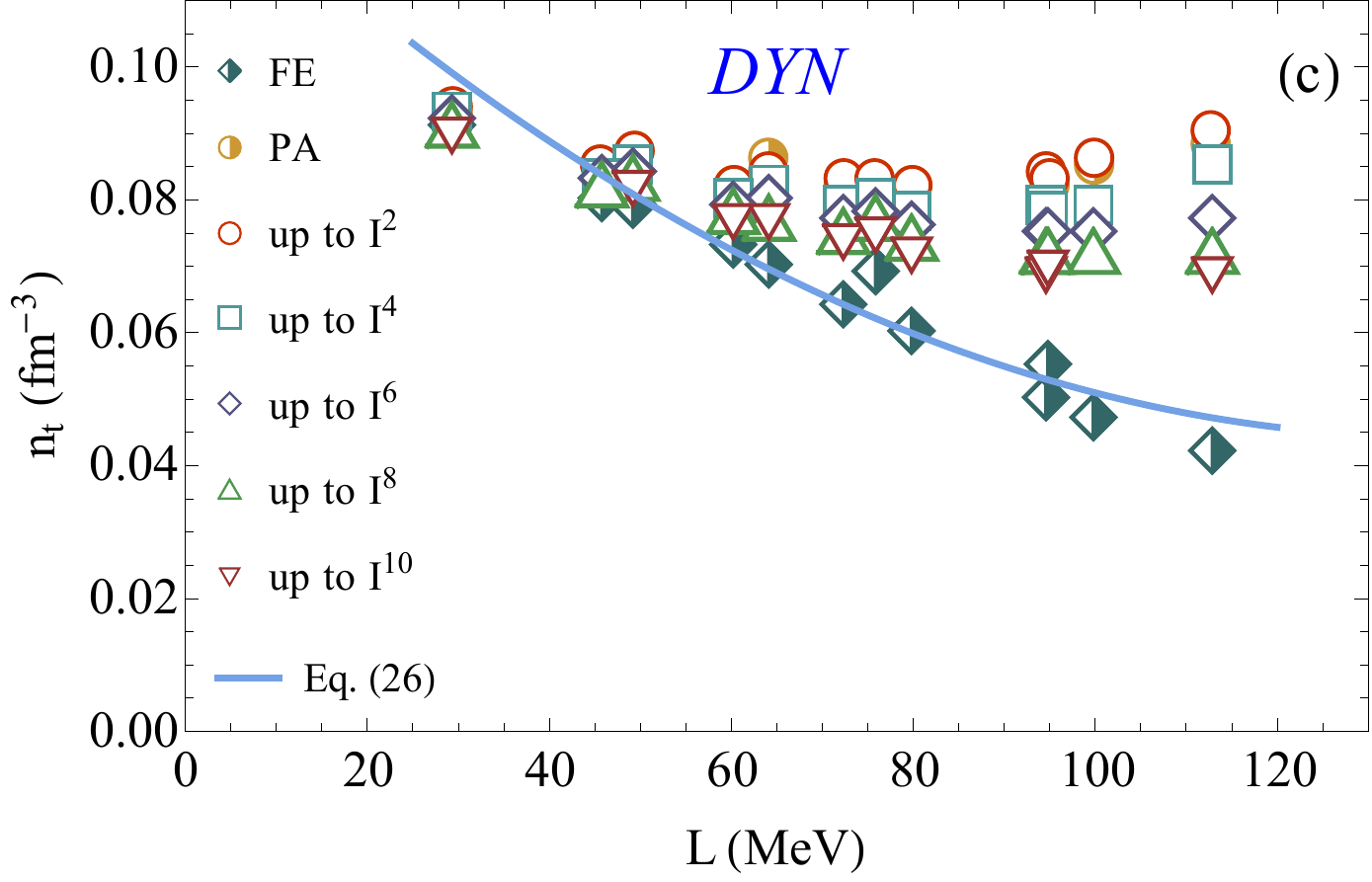}
	\hspace{1cm}
	\includegraphics[width=7.6cm, height=5.6cm]{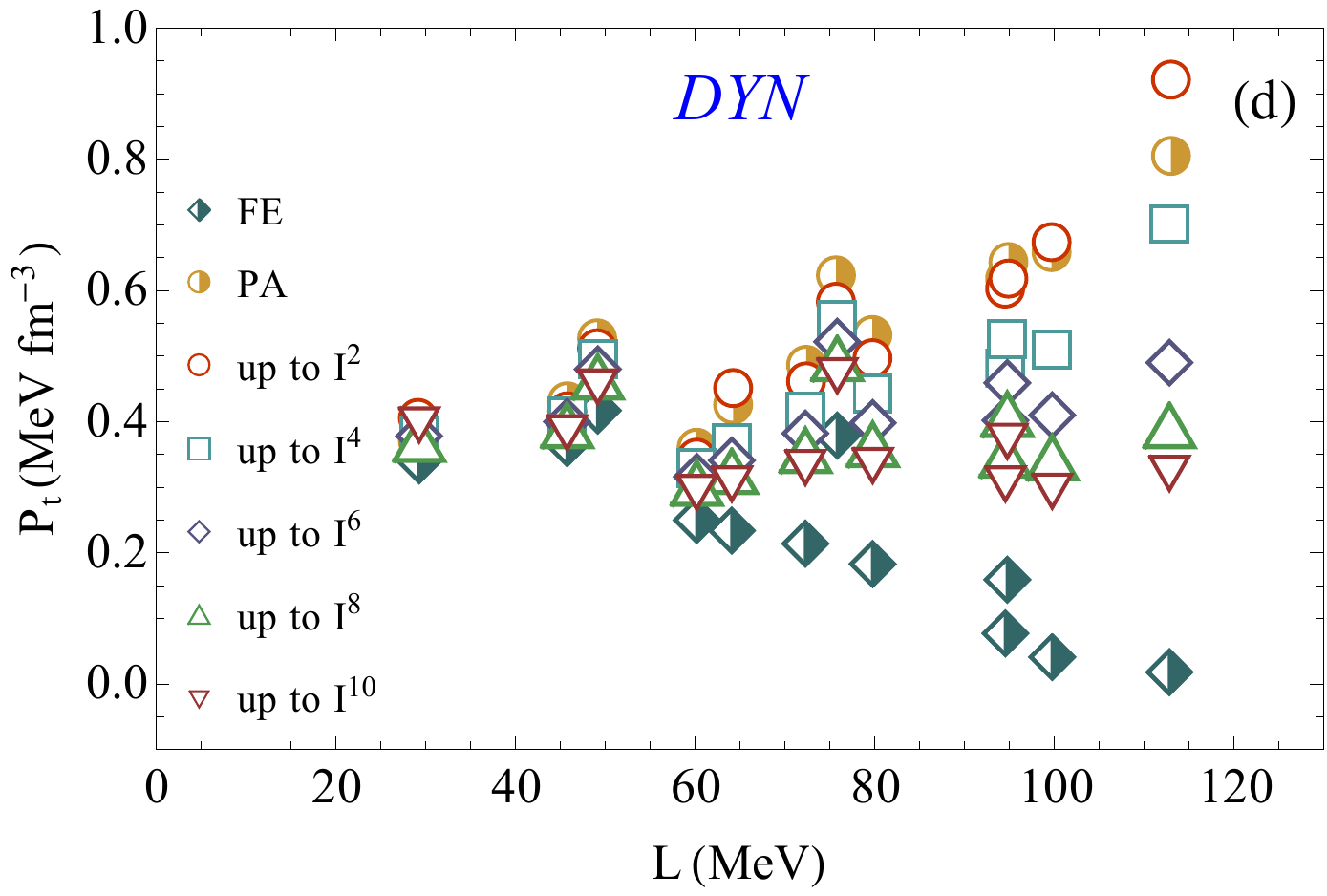}
	\caption{Transition density $n_{\rm t}$ and pressure $P_{\rm t}$ as a function of the slope parameter $L$ for various nuclear models. The calculations are performed with the thermodynamical method for (a,b) and the dynamical one for (c,d). The half-filled diamonds present the full expression; the half-filled circles present the parabolic approximation; and the circles, the squares, the diamonds, the triangles, and the reversed triangles present the approximation up to second, fourth, sixth, eighth, and tenth order, respectively, based on Eq.~\eqref{esym-exp-1}. The solid line in (a,c) represents the Eq.~\eqref{n_t-L}.} 
	\label{fig:tr_n_p}
\end{figure*}

In the present paper, we compare a class of EoSs generated from three different nuclear models, as shown in Tables~\ref{tab:1} and~\ref{tab:2}. In particular, we employ a momentum-dependent interaction model (MDI) which was presented and analyzed in previous papers~\cite{Prakash-97,Moustakidis-08}. The parametrization of the model has been chosen in order to generate specific values for $E_{\rm sym,2}(n_{s})$ and the slope parameter $L$ at the saturation density $n_s$. The second one, that is the HLPS model, is based on the microscopic calculations, in the framework of chiral effective field theory interactions, in low densities and suitable polytropic parametrization at high nuclear densities~\cite{Hebeler-13}. Specifically, we employ two parametrizations of the model, the soft ($L=29.5~{\rm MeV}$) and the stiff ($L=49.4~{\rm MeV}$) ones (for more details see Ref.~\cite{Hebeler-13}). Lastly, we use three versions of the Skyrme model that is the SkI4, Ska and Sly4~\cite{Chabanat-97,Farine-97}. The above nuclear models are used to construct the EoS of the core of a neutron star. The EoS of the crust is taken from the well known model of Baym, Pethick, and Sutherland~\cite{Baym-71b} (hereafter the BPS model). Although, in the present paper we mainly focus on the crust-core transition effects, the models have been chosen in order to produce, even marginally,  the limit of the two solar masses.

In Tables~\ref{tab:1} and~\ref{tab:2} we present our calculations for the transition densities and the corresponding pressures. Inside the parentheses is the value of the slope parameter $L$. Also, the specific cases MDI(95-30) and MDI(95-32) correspond to $L=95$ MeV and $E_{\rm sym,2}(n_s)=30$ and $32~{\rm MeV}$, respectively. The calculations are performed using both the thermodynamical and the dynamical method. In each case, we employed the full expression of the energy per baryon of each model, the parabolic approximation [see Eq.~\eqref{E-pa}], and the corresponding terms of the expansion [see Eq.~\eqref{esym-exp-1}] up to tenth order. In order to clarify further the predictions, we display the results also in Fig.~\ref{fig:tr_n_p}.

In each case, the higher the order in the expansion, the lower the values of $n_{\rm t}$ and $P_{\rm t}$. Even more, the higher the value of the slope parameter $L$, the larger the deviation between the second order predictions and the consideration of the full expression. In other words, according to our finding, the lower the value of $L$, the higher the accuracy of the parabolic approximation.  It is worth mentioning here that the quadratic dependence of $P_{\rm t}$ on $n_{\rm t}$ (see Eq.~\eqref{Pr-tra}) is well reflected in the current predictions and mainly in the dispersion of the results for high values of $L$. As a general rule, the thermodynamical method leads to higher values of $P_{\rm t}$ on $n_{\rm t}$ compared to the dynamical one. However, the most distinctive feature is the appearing of a universal dependence of $n_{\rm t}$ on $L$ in both methods concerning the full expression. We found that, independently of the employed model, there is an ordering on the mentioned dependence, where the increase of $L$, leads to a decreased $n_{\rm t}$. In order to check our finding, we utilize the following expression, predicted by Steiner \textit{et al.}~\cite{Steiner-2015}:
\begin{equation}
    \hspace{-0.2cm}
	n_t=S_{30}\left(0.1327-0.0898 L_{70}+0.0228 L_{70}^2\right) \ ({\rm fm}^{-3}),
	\label{n_t-L}
\end{equation}
where $S_{30}=E_{\rm sym,2}(n_s)/(30~{\rm MeV})$ and $L_{70}=L/(70~{\rm MeV})$. We found that there is an excellent agreement with the present results [see Figs.~\ref{fig:tr_n_p}(a) and~\ref{fig:tr_n_p}(c)], concerning the dynamical method, for a large interval of $L$ and  mainly for different nuclear models.

\begin{figure*}
	\centering
	\includegraphics[width=\textwidth]{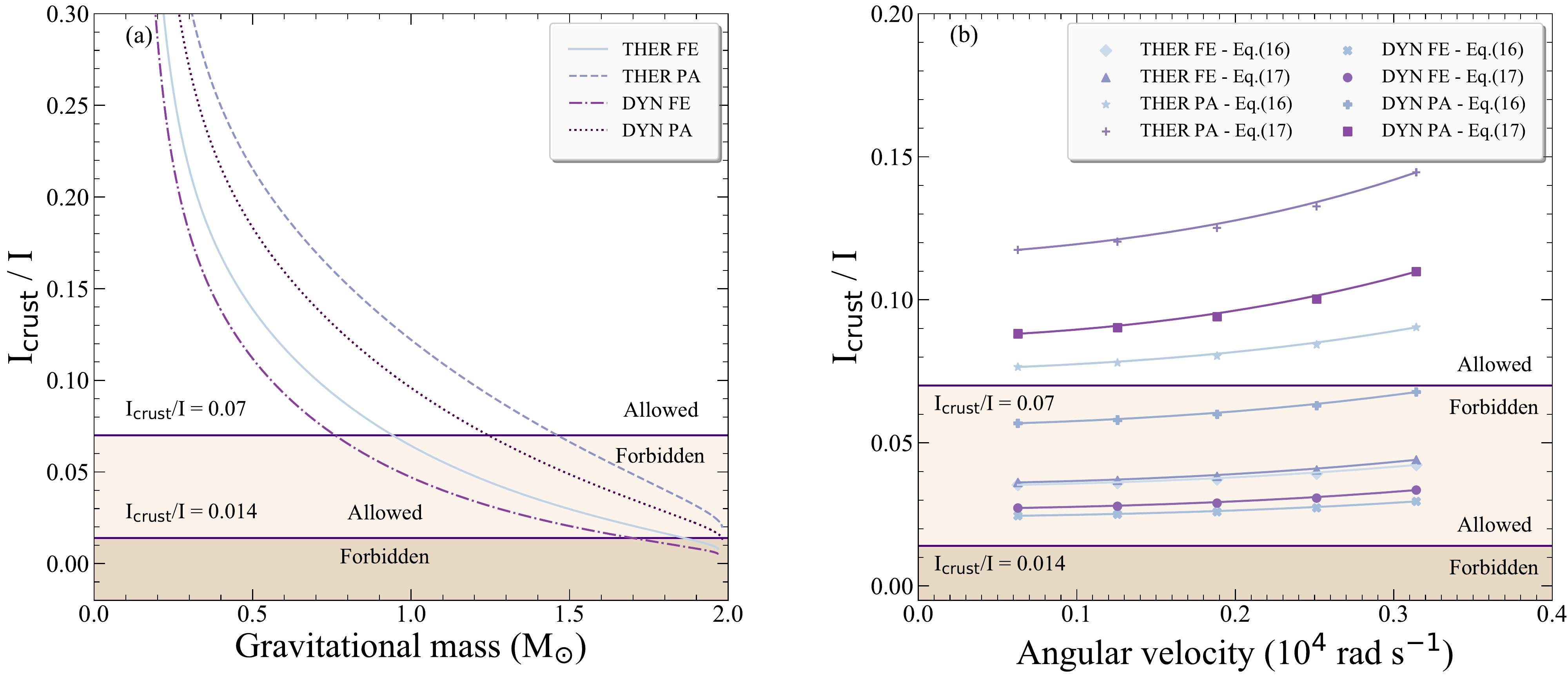}
	\caption{(a) The crustal fraction of moment of inertia as a function of the gravitational mass for the four methods using the approximation~\eqref{crust-1} and the CIF matching process. (b) The crustal fraction of moment of inertia for a  neutron star with $M = 1.4~M_{\odot}$ as a function of the angular velocity for the four methods, using the approximations~\eqref{crust-1} and~\eqref{crust-2}, and the CIF matching process. The regions defined by $I_{\rm crust}/I=0.07$ and $0.014$, which represent a possible constraint deduced for the Vela pulsar assuming a neutron star with $M = 1.4~M_{\odot}$, are also included (for more details see text).} 
	\label{fig:crustal-inertia}
\end{figure*}

In order to enrich the discussion about the effect of the momentum dependence on the higher order of the expansion of the energy, it is worth mentioning that for the case of the Gogny forces it was found that the zero-range, the direct (density-dependent), and the exchange (momentum-dependent) terms contribute with similar magnitudes to the determination of
$E_{\rm sym,2}$ (see also Refs.~\cite{Boquera-2017,Sellahewa-2014}). However, their contribution is with
different signs, which leads to cancellations in $E_{sym,2}$ between the power-law zero-range term, the linear density-dependent direct term, and the more complex exchange term. Depending on the parametrization, the sum of the zero-range and direct terms is positive whereas the exchange term is negative, and vice versa. In any case, there is a balance among the terms, which gives rise to a similar density dependence of the symmetry energy coefficient $E_{\rm sym,2}$ for all parameter sets. Nevertheless, this is not the case for higher-order terms because both the zero-range and the direct components of the energy per particle depend on the quadratic term of the isospin asymmetry. In other words, the higher order corrections to the symmetry energy are only sensitive to the kinetic term and to the momentum-dependent term, i.e. the exchange term of the Gogny force. We note that the same
pattern is found in zero-range Skyrme forces. In addition, in Skyrme forces, the higher-order symmetry energy coefficients arise exclusively from the kinetic and  momentum-dependent terms of the interaction (for a detailed discussion see Refs.~\cite{Boquera-2017,Sellahewa-2014}).

Moreover, we studied the effects of the crust-core interface on four specific properties of neutron stars. We employed in each case, as an example, the MDI(80) model. It has to be noted here that we found similar results for the rest of the models (depending of course on the value of the slope parameter $L$). The matching at the interface, between the crust and the core EoSs, is achieved  by employing two methods. In the first case, which is the code-interpolating fitting (hereafter CIF), the matching is performed automatically by the employed PYTHON code. In the second case, which is the thermodynamically consistent one (hereafter TC$_i, \ i=1,2,3$), the matching is performed with respect to the thermodynamic consistency following  the recipe given in Ref.~\cite{Fortin-2016} (and references therein; more details are given in Appendix).

\subsection{Crustal fraction}
The effects of the symmetry energy on the crustal fraction of a nonrotating and slowly rotating with $M=1.4~M_{\odot}$ neutron star using the CIF matching process are displayed in Fig.~\ref{fig:crustal-inertia}. In particular, we employed the  approximations~\eqref{crust-1} and~\eqref{crust-2} for comparison.  Our finding confirms previous predictions for a nonrotating neutron star; that is, the crustal moment of inertia is very sensitive to the location of the crust-core interface. In particular, in Fig.~\ref{fig:crustal-inertia}(a), we display the dependence of the crustal ratio $I_{\rm crust}/I$ on the gravitational mass. The application of the dynamical method, using the full expression, leads to the lower values of $I_{\rm crust}/I$, compared to the thermodynamical one. The two horizontal lines represent, each one, a possible constraint on $I_{\rm crust}/I$ deduced for the Vela pulsar (assuming a neutron star with $M = 1.4~M_{\odot}$). The lower limit, 0.014, was suggested in Ref.~\cite{Link-99}, while the higher one, $0.07$, was considered in Refs.~\cite{Andersson-2012,Chamel-2012} in order to explain the glitches. It is notable, in Fig.~\ref{fig:crustal-inertia}(b), that the predictions of the approximations~\eqref{crust-1} and~\eqref{crust-2} are almost identical in the case of the full expression (in both methods), while distinct deviations appear in the case of the parabolic approximation. 

Moreover, in order to further clarify the effects of the matching process on the relevant predictions, we present in Table~\ref{tab-rad} the corresponding values of the radius $R_{1.4}$ and the crustal fraction $(I_{\rm crust}/I)_{1.4}$ at $1.4~M_{\odot}$ configuration. As a general rule, the CIF matching leads to higher values of the radius compared to the TC$_{i}$ cases. It is noteworthy that the TC$_{i}$ matching processes give similar values. Our finding leads to the conclusion that the variation of the values due to the applied method is about twice as high as the variation due to the matching process (approximately $0.75-0.85 \%$ due to the method and $0.46-0.47\%$ due to the matching process). However, the deviations presented due to the specific choice of the method are more dramatic in the case of the crustal fraction compared to the matching processes.

\begin{figure*}
	\centering
	\includegraphics[width=\textwidth]{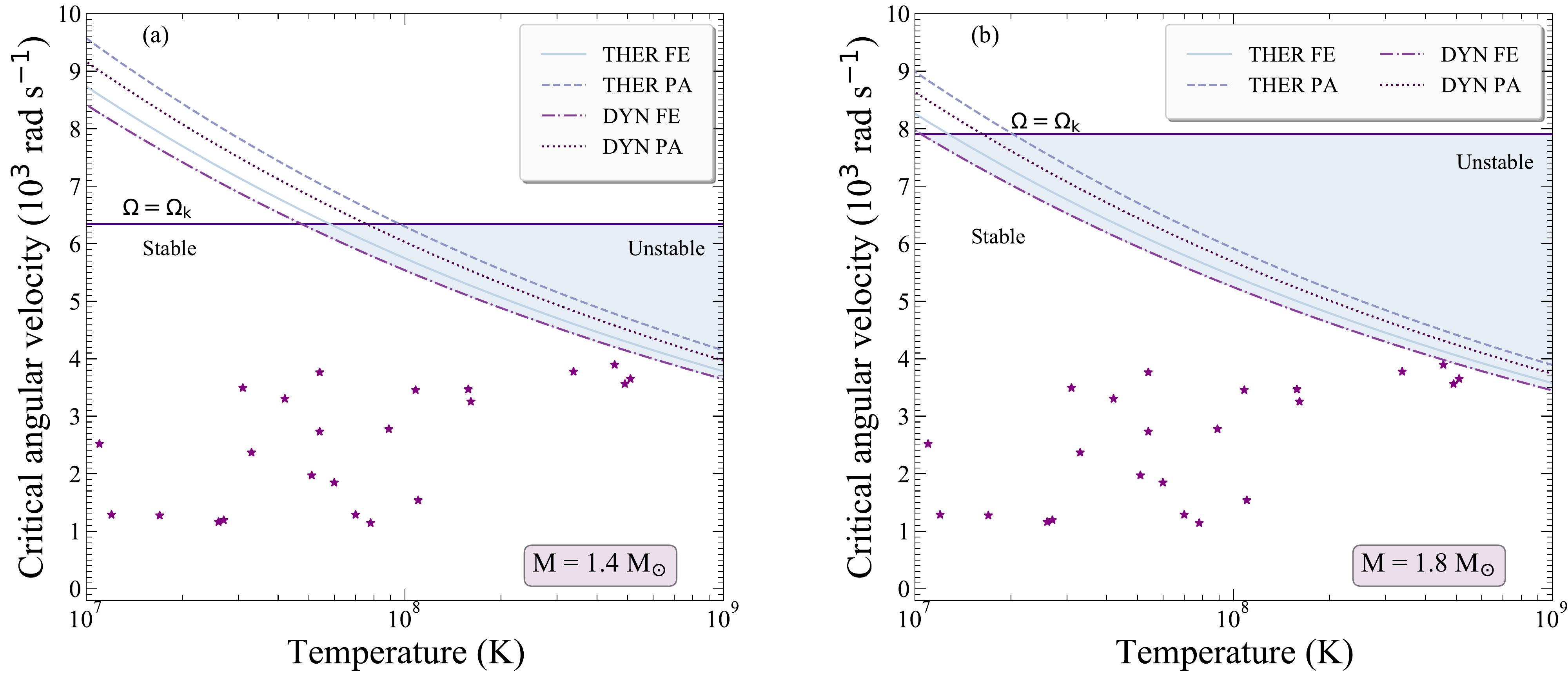}
	\caption{The critical angular velocity as a function of the temperature for the four methods for a neutron star with (a) $M=1.4~M_{\odot}$ and (b) $M=1.8~M_{\odot}$ using the CIF matching process. For comparison, the observed cases of LMXBs and MSRPs from Haskell \textit{et al.}~\cite{Haskell-2012} presented with stars and the horizontal lines, which correspond to the Kepler angular velocity $\Omega_{\rm k}$~\cite{Koliogiannis-2020}, are also included. The shaded region corresponds to the r-mode instability window.} 
	\label{fig:r-mode-1.4-1.8}
\end{figure*}

\begin{table*}
	\caption{The radius and the fraction $I_{\rm crust}/I$ of nonrotating neutron stars using the MDI(80) nuclear model for the core and the BPS model~\cite{Baym-71b} for the crust at the $M_{\rm gr}=1.4~M_{\odot}$ configuration. The data correspond to the four methods and matching processes.}
	\label{tab-rad}
	\begin{ruledtabular}
		\begin{tabular}{l|cccccccc}
			 & \multicolumn{4}{c}{$R_{1.4}$ (km)} & \multicolumn{4}{c}{$\left(I_{\rm crust}/I\right)_{1.4}$} \\
			 \cline{2-5}\cline{6-9}
			Nuclear model & CIF & TC$_{1}$ & TC$_{2}$ & TC$_{3}$ & CIF & TC$_{1}$ & TC$_{2}$ & TC$_{3}$ \\
			\hline
			THER FE & 12.973 & 12.918 & 12.911 & 12.915 & 0.035 & 0.034 & 0.034 & 0.034 \\
			THER PA & 13.050 & 12.985 & 12.993 & 12.986 & 0.076 & 0.075 & 0.075 & 0.075 \\
			DYN FE & 12.958 & 12.911 & 12.900 & 12.912 & 0.024 & 0.024 & 0.024 & 0.024 \\
			DYN PA & 13.071 & 13.010 & 13.010 & 13.004 & 0.056 & 0.055 & 0.055 & 0.055 \\ 
		\end{tabular}
	\end{ruledtabular}
\end{table*}
\begin{table}
	\caption{The fraction $I_{\rm crust}/I$ of rotating neutron stars using the MDI(80) nuclear model for the core and the BPS model~\cite{Baym-71b} for the crust at the $M_{\rm gr}=1.4~M_{\odot}$ configuration and $\Omega = 3\times 10^{3}~{\rm rad~s^{-1}}$. The top and bottom panels correspond to the approximations~\eqref{crust-1} and~\eqref{crust-2}, respectively. The data correspond to the four methods and matching processes.}
	\label{tab-diff}
	\begin{ruledtabular}
		\begin{tabular}{l|cccc}
			Nuclear model & CIF & TC$_{1}$ & TC$_{2}$ & TC$_{3}$ \\ 
			\hline
			THER FE & 0.042 & 0.041 & 0.041 & 0.041 \\
			THER PA & 0.089 & 0.087 & 0.088 & 0.087 \\
			DYN FE & 0.029 & 0.029 & 0.028 & 0.029 \\
			DYN PA & 0.067 & 0.065 & 0.065 & 0.065 \\
			\newline \\
			THER FE & 0.043 & 0.043 & 0.042 & 0.043 \\
			THER PA & 0.142 & 0.139 & 0.139 & 0.139 \\
			DYN FE & 0.033 & 0.033 & 0.032 & 0.033 \\
			DYN PA & 0.108 & 0.105 & 0.105 & 0.105 \\
		\end{tabular}
	\end{ruledtabular}
\end{table}

Furthermore, Table~\ref{tab-diff} displays the crustal fraction $(I_{\rm crust}/I)_{1.4}$ at $\rm \Omega = 3\times 10^{3}~{\rm rad~s^{-1}}$ for the various methods and matching processes. We note that the results lead to an error estimation of $\le 2\%$ for both approximations (\ref{crust-1},~\ref{crust-2}) between the methods and the matching processes. In this case too, the deviations between the parabolic approximations (dynamical and thermodynamical), and also between the parabolic approximation and the full expression, are significant. The latter has its origin in the values of the radius, which seems to play an important role. The results confirm the role played by the correct location of the transition density on the crustal moment of inertia, at least on static and slowly rotating neutron stars.

We conjecture that in the case of a rapidly rotating neutron star (close to the mass-shedding limit, that is the Kepler angular velocity) the effects of the symmetry energy expansion on $I_{\rm crust}/I$ will be dramatic. In this case, one must carefully select the appropriate method with the full expression. Otherwise, the accuracy of the  predictions will suffer from large uncertainties.

\subsection{$r$-mode instability}
\begin{figure*}
	\includegraphics[width=\textwidth]{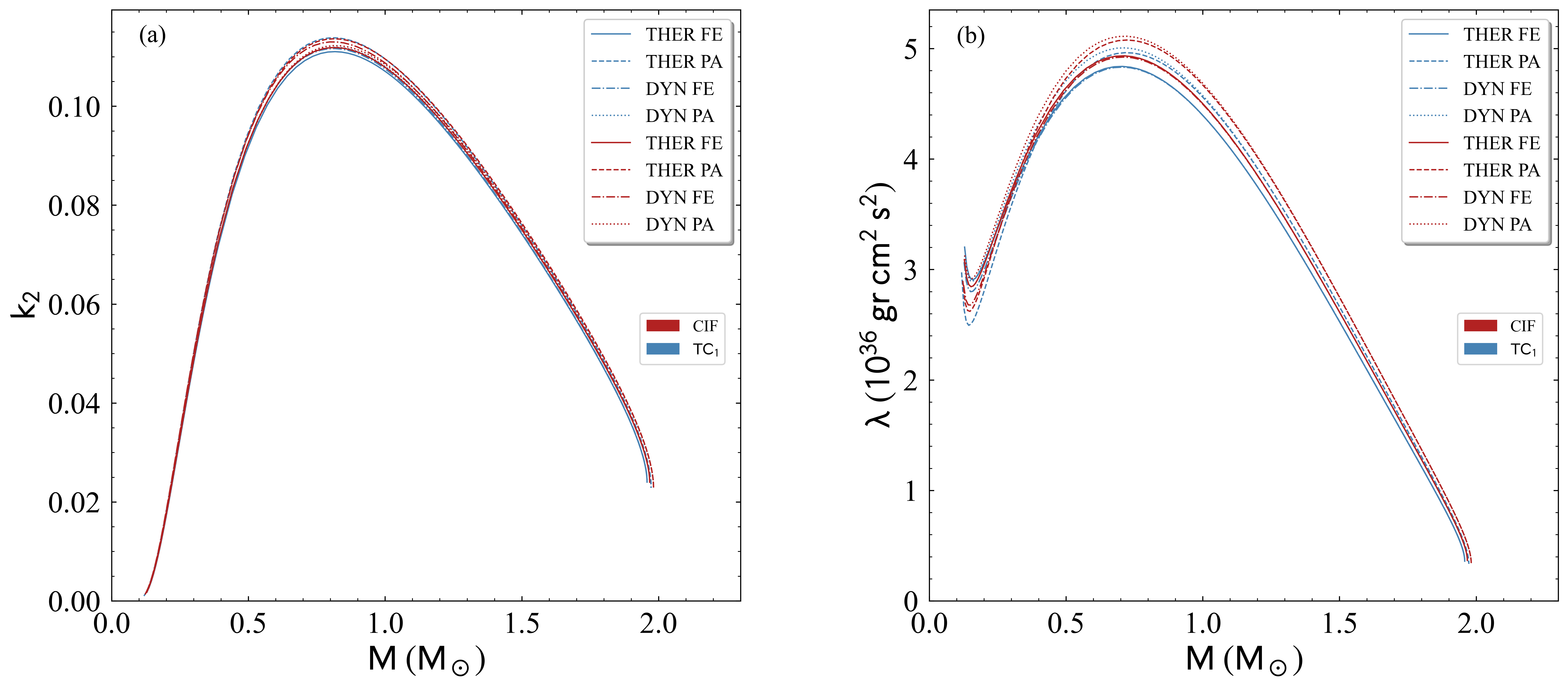}
	\caption{(a) The tidal Love number $k_2$ and (b) the tidal deformability $\lambda$ as a function of the gravitational mass for the four methods and the matching processes CIF and TC$_1$.} 
	\label{fig:k2}
\end{figure*}

In the present paper we concentrated also on the effects of the crust-core transition on the $r$-mode instabilities. According to Eq.~\eqref{r-mode-3}, the critical angular velocity $\Omega_{c}$ is sensitive to the neutron star core radius $R_{\rm core}$ and energy density ${\cal E}_{\rm t}$. In Fig.~\ref{fig:r-mode-1.4-1.8} we display the $r$-mode instability windows for the selected four cases (MDI model with $L=80~{\rm MeV})$ and for $M=1.4~M_{\odot}$ and $1.8~M_{\odot}$ using the CIF matching process, respectively. Moreover, we included many cases of LMXBs and a few millisecond radio pulsars (MSRPs). It is worth mentioning that the estimates of core temperature $T$ have large uncertainties (see Ref.~\cite{Haskell-2012}). However, since the purpose of the present paper is to exhibit the role of the crust-core transition, more details on the temperature uncertainties are not included.

According to the relevant Figures, the full expression cases (dynamical and thermodynamical) lead to lower values compared to the corresponding parabolic approximations, increasing the instability window. In other words, the parabolic approximation gives rise to a narrower instability window, for the same value of temperature. Therefore, since the damping due to viscous dissipation at the boundary layer of the perfectly rigid crust and fluid core is of major importance in $r$-mode studies, one has to use carefully the employed method for the estimation of the crust-core edge. The role played by the method and the matching process is clarified in Table~\ref{tab-diff_1}. In particular, we present the values of  the critical angular velocity $\Omega_{c}$ for a neutron star with mass $1.4 M_{\odot}$ at temperature $T=10^8$ K for the various methods and matching processes. We found that the effect of the method is much more dramatic (deviations around $12 \%$) compared to the effect of the matching process which leads to a deviation less than $0.3\%$. Our findings confirm the reliability of the predictions concerning the critical angular velocity.
\begin{table}
	\caption{The critical angular velocity of neutron stars in units of $10^{3}~{\rm rad~s^{-1}}$ using the MDI(80) nuclear model for the core and the BPS model~\cite{Baym-71b} for the crust at temperature of $10^{8}~{\rm K}$. The data correspond to the four methods and matching processes.}
	\label{tab-diff_1}
	\begin{ruledtabular}
		\begin{tabular}{l|cccc}
			Nuclear model & CIF & TC$_{1}$ & TC$_{2}$ & TC$_{3}$\\ 
			\hline
			THER FE & 5.746 & 5.763 & 5.759 & 5.758 \\
			THER PA & 6.298 & 6.282 & 6.286 & 6.283 \\
			DYN FE & 5.538 & 5.555 & 5.549 & 5.547 \\
			DYN PA & 6.030 & 6.036 & 6.035 & 6.035 \\ 
		\end{tabular}
	\end{ruledtabular}
\end{table}
%

\subsection{Tidal deformability}
%
\begin{figure*}
	\includegraphics[width=\textwidth]{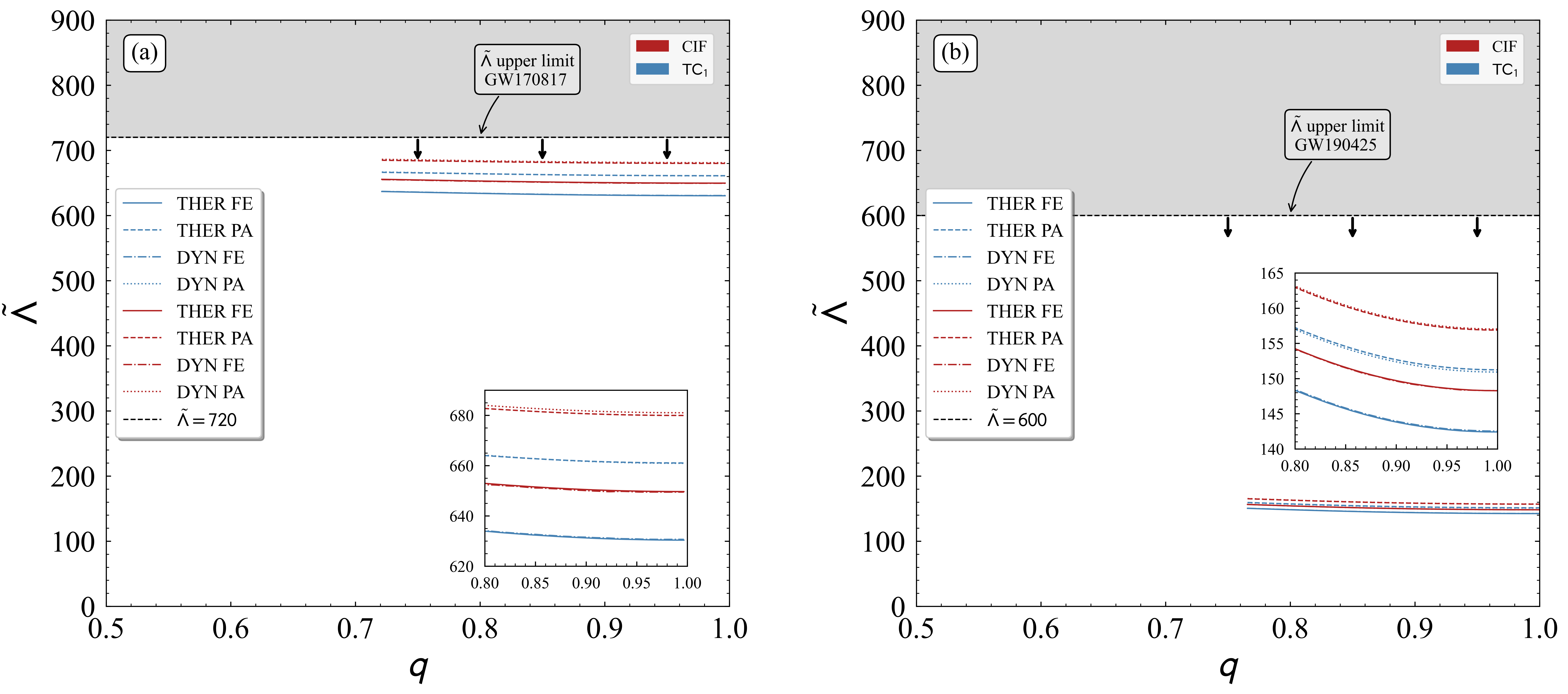}
	\caption{The effective tidal deformability $\tilde{\Lambda}$ for (a) the GW170817~\cite{Abbott-2018,Abbott-2019} event and (b) the GW190425~\cite{Abbott-2020} event, as a function of the binary mass ratio $q$ for the four methods and the CIF and TC$_{1}$ matching processes. The horizontal black lines indicate the observed upper limit on $\tilde{\Lambda}$, and the gray shaded region corresponds to the excluded values for $\tilde{\Lambda}$. The inner figures visualize the behavior of the CIF and the TC$_{1}$ matching processes in the area under consideration.} 
	\label{fig:Ltilde}
\end{figure*}
The effects of the crust-core interface may also be studied through the tidal deformability, as derived from gravitational wave events. We consider the two recent events, that is, GW170817~\cite{Abbott-2018,Abbott-2019} and GW190425~\cite{Abbott-2020}. The chirp masses for the two events are ${\cal M}_c=1.186\ M_{\odot}$ and $1.144\ M_{\odot}$, respectively. In addition, according to data from LIGO, the component masses vary in the ranges $m_1 \in (1.36,1.60)\ M_{\odot}$ and $m_2 \in (1.16,1.36)\ M_{\odot}$, and $m_1 \in (1.654,1.894)\ M_{\odot}$ and $m_2 \in (1.45,1.654)\ M_{\odot}$, respectively. It has to be noted that we have modified the range of the component masses to have an equal-mass boundary, i.e. the binary mass ratio, to be $q\leq1$.

In Fig.~\ref{fig:k2} we display the Love number $k_2$ as a function of the gravitational mass for the various methods and the matching processes CIF and TC$_{1}$. Obviously, $k_2$ is almost insensitive to the approach. However, in the case of the tidal deformability $\lambda$, the effects are more pronounced especially for low neutron star masses. In particular, in the case of full expression, both approaches lead to similar predictions, while in the parabolic approximation both cases lead to higher values of $\lambda$. In general, the matching process TC$_{1}$ (blue color) shifts the curves downwards in comparison to the CIF matching process (red color). We notice that the effects among the methods and matching processes are small but not negligible (see also the recent Ref.\cite{Passamonti-2021}). 

Moving on to the study of the two observed events for binary neutron star mergers (GW170817 and GW190425), the results are displayed in Fig.~\ref{fig:Ltilde}. One can observe that the CIF and TC$_1$ matching processes for full expressions and parabolic approximations in both events lead to an almost identical behavior. The main difference is between the full expression and parabolic approximation, with the latter leading to higher values of $\tilde{\Lambda}$. Moreover, the matching process TC$_1$ shifts downwards the curves in comparison to the CIF matching process. This behavior is present in both events and the explanation lies in the dependence of the tidal deformability on the radius [see Eq.~\eqref{Lambda-dim}]. More specifically, Tables~\ref{tab:1} and~\ref{tab:2} show that the parabolic approximation corresponds to higher values of the transition density than the full expression and, consequently, to higher values of the radius. The latter conclusion is based on the $1.4~M_{\odot}$ configuration from Table~\ref{tab-rad}. It is also noteworthy that the CIF matching process provides higher values of the radius than the TC$_{i}$ one for all methods. Therefore the shift of the curves to lower values of $\tilde{\Lambda}$ in the matching process TC$_1$ is expected.

Additionally, we notice that for the second event (GW190425) the curves are shifted to much lower values of $\tilde{\Lambda}$ compared to the GW170817 event. This kind of behavior is based on the fact that the component masses and the chirp mass of the second event are higher than the first one. Also, the differences between the curves in the second event are smaller. Since the crust affects low and intermediate masses more than the high ones, the effects of the different methods on the crust-core transition are more evident in events with lower masses. We postulate that for the transition density range of our paper ($n_{\mathrm {t}}<0.1\;\mathrm{fm^{-3}}$) and for events with lower component masses (equivalently lower chirp mass $\mathcal{M}_c$), such as GW170817, the role of the crust-core transition may be significant in the EoS. The latter can lead to the exclusion or acceptance of an approach with respect to the observed upper limit of $\tilde{\Lambda}$.

\begin{figure*}
	\includegraphics[width=\textwidth]{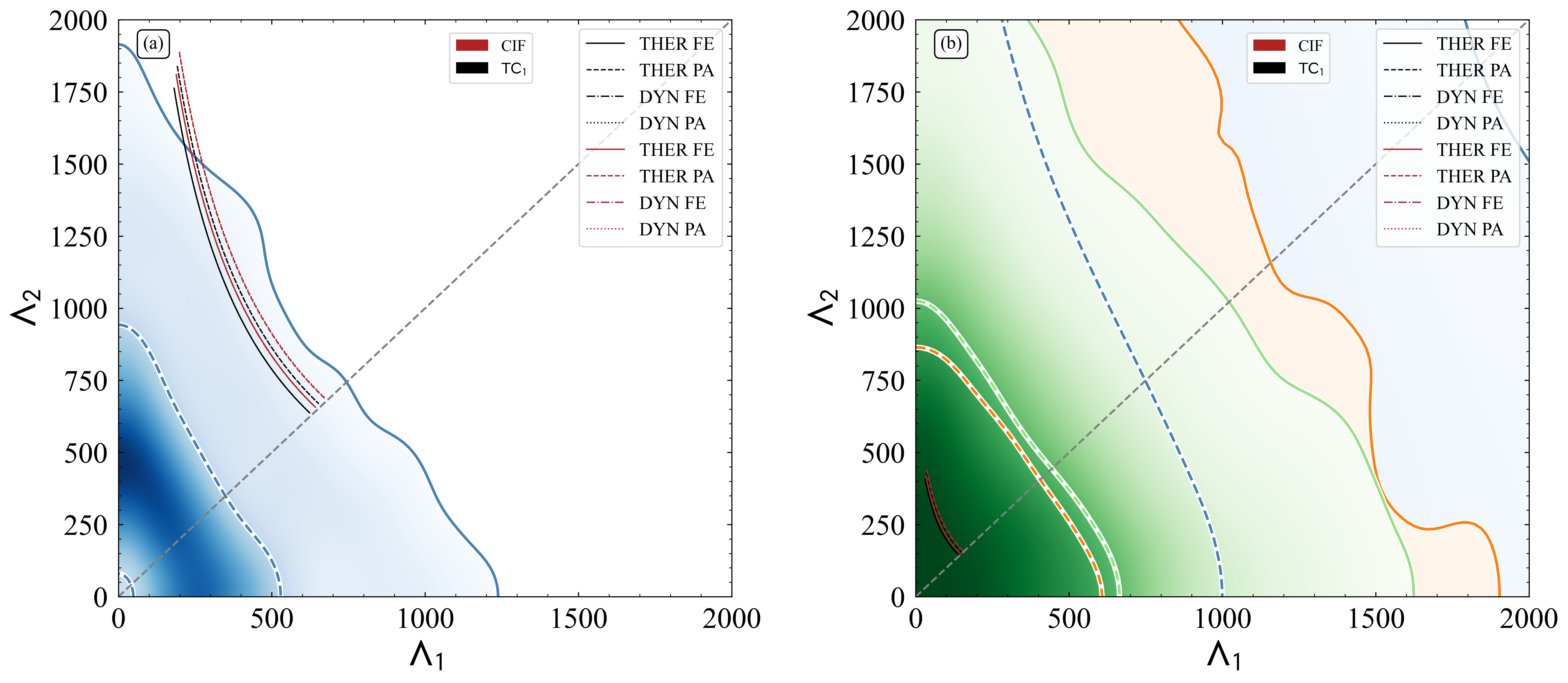}
	\caption{The dimensionless tidal deformability of the lighter component neutron star $\Lambda_2$ for (a) the GW170817~\cite{Abbott-2018,Abbott-2019} event and (b) the GW190425~\cite{Abbott-2020} event, as a function of the dimensionless tidal deformability of the heavier  component neutron star $\Lambda_1$, for the four methods and the CIF and TC$_{1}$ matching processes. The red (black) curves correspond to the CIF (TC$_1$) matching process. The shaded regions indicate the estimating posterior distribution for both events (see Refs.~\cite{Abbott-2018,Abbott-2019,Abbott-2020}). The dashed colored (solid) line indicates the $50\%$ ($90\%$) credible region. The diagonal dashed line corresponds to $\Lambda_1=\Lambda_2$.}
	\label{fig:L1L2}
\end{figure*}

In Fig.~\ref{fig:L1L2} we display the $\Lambda_2-\Lambda_1$ space for both events and the CIF and TC$_{1}$ matching processes. The behavior of the curves is similar to Fig.~\ref{fig:Ltilde}. In the second event with higher masses, the curves are shifted to lower values of $\Lambda$, and are less distinguishable compared to the first event in Fig.~\ref{fig:L1L2}(a). We notice that in Fig.~\ref{fig:L1L2}(b) the shaded regions indicate three different posterior distributions (derived from three different waveform models) for the GW190425; the displayed upper limit contours correspond to the constrained value 1200 for $\tilde{\Lambda}$~\cite{Abbott-2020}.
%

\subsection{ Minimum mass of neutron star}
\begin{table*}
	\caption{The minimum mass configuration (including the minimum mass $M_{\rm min}$ in units of $M_{\odot}$, the corresponding radius $R_{\min}$ in units of km, and the central energy density $\rho_{c}$ in units of $10^{14}~{\rm gr~cm^{-3}}$) of nonrotating neutron stars using the MDI(80) nuclear model for the core and the BPS model~\cite{Baym-71b} for the crust. The data correspond to the four methods and matching processes.}
	\label{tab-Mmin}
	\begin{ruledtabular}
		\begin{tabular}{l|ccccccccccccc}
			 & \multicolumn{4}{c}{$M_{\rm min}$} & \multicolumn{4}{c}{$R_{\rm min}$} & \multicolumn{4}{c}{$\rho_{c}$}\\
			 \cline{2-5}\cline{6-9}\cline{10-13}
			Nuclear model & CIF & TC$_{1}$ & TC$_{2}$ & TC$_{3}$ & CIF & TC$_{1}$ & TC$_{2}$ & TC$_{3}$ & CIF & TC$_{1}$ & TC$_{2}$ & TC$_{3}$ \\ 
			\hline
			THER FE & 0.0920 & 0.0930 & 0.0919 & 0.0925 & 245 & 247 & 242 & 244 & 2.135 & 2.131 & 2.153 & 2.142 \\
			THER PA & 0.0919 & 0.0920 & 0.0919 & 0.0921 & 219 & 222 & 220 & 228 & 2.141 & 2.160 & 2.154 & 2.155 \\
			DYN FE & 0.0896 & 0.0918 & 0.0894 & 0.0921 & 255 & 243 & 254 & 241 & 2.154 & 2.153 & 2.172 & 2.151 \\
			DYN PA & 0.0922 & 0.0922 & 0.0922 & 0.0921 & 243 & 241 & 244 & 235 & 2.109 & 2.123 & 2.123 & 2.134 \\ 
		\end{tabular}
	\end{ruledtabular}
\end{table*}

\begin{figure}
	\includegraphics[width=0.5\textwidth]{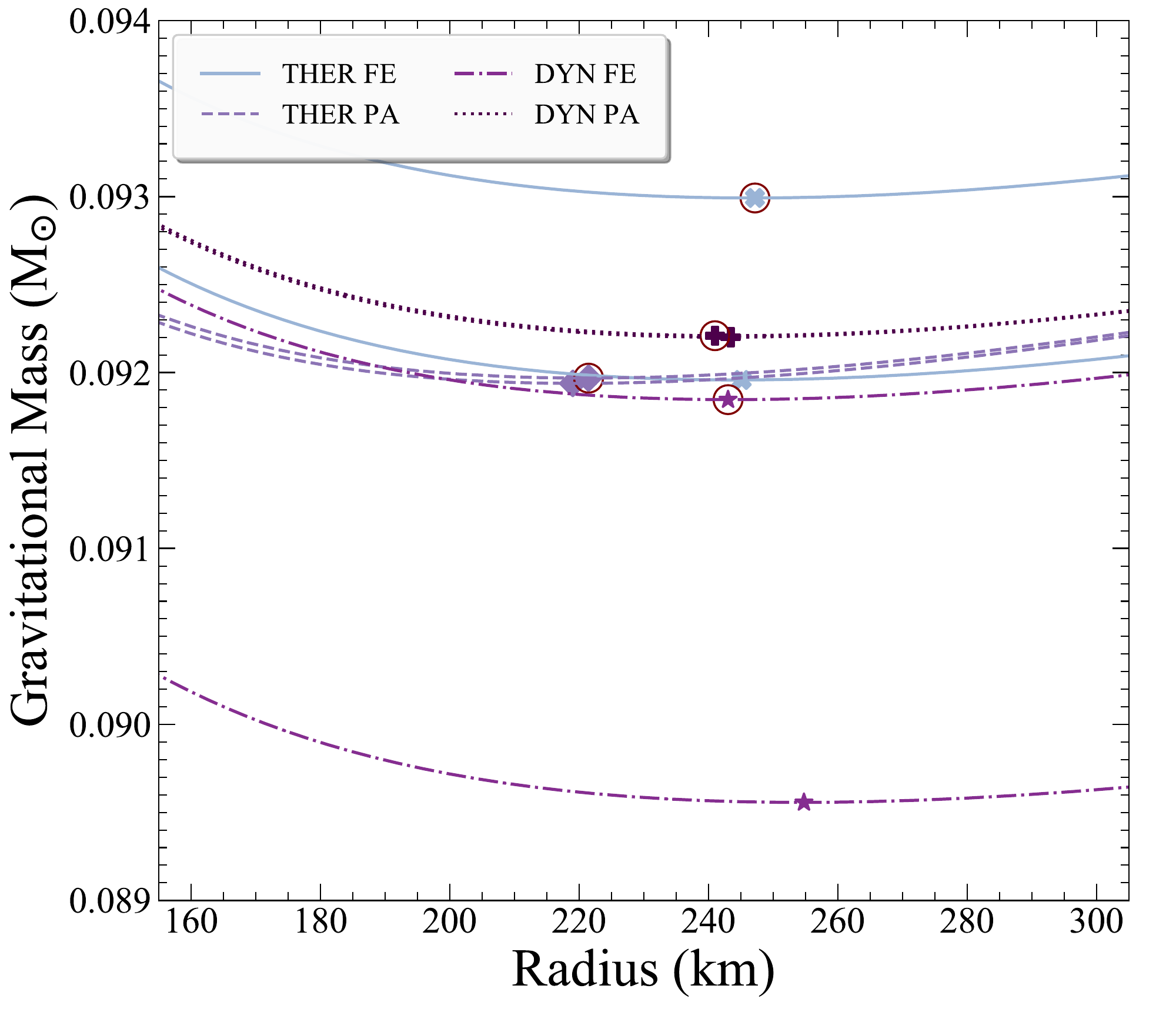}
	\caption{The gravitational mass as a function of the radius near the minimum gravitational mass region, for the four methods and the CIF and TC$_{1}$ matching processes. In the case of the CIF matching process, the minimum mass configuration is indicated with a cross for the thermodynamical FE method, a diamond for the thermodynamical PA method, a star for the dynamical FE method, and a plus sign for the dynamical PA method. In the TC$_{1}$ case the results are indicated with the corresponding circles.}
	\label{fig:min_mass}
\end{figure}

In addition, in Table~\ref{tab-Mmin} we present the minimum mass, the corresponding radius, and central energy density for the various methods and matching processes. We expect that the results must be very sensitive to the matching process. To be more specific, it should be emphasized that an accurate treatment of the minimum mass demands the use of the same nuclear model both for the core and for the crust region (and consequently on the crust-core interface). However, since our main purpose was to examine the effect of the symmetry energy on the location of the transition density and pressure, we used the same nuclear model for the core while the EoS of the crust is taken from the well known BPS model~\cite{Baym-71b}. Since, this method may suffer from an expected uncertainty, we tried to treat it by employing the various matching processes. Although our procedure does not ensure the high accuracy of the results, in each case useful insight may be gained. Moreover, the extent of the sensitivity of the minimum mass configuration to the transition density and to the employed matching process may be revealed.
In Fig.~\ref{fig:min_mass} we display the gravitational mass as a function of the radius near the minimum gravitational mass region using the CIF and TC$_{1}$ matching processes.
Our predictions are close to those found by Haensel \textit{et al.}~\cite{Haensel-2002} using two different nuclear models. Obviously, as expected, the effects on the minimum mass are almost negligible. However, the effects are more pronounced in the case of the corresponding radius, depending both on the approach and on the expansion. In particular, taking into consideration the most accurate method, that is, the full expression in dynamical method, and the least accurate one, that is, the parabolic approximation in thermodynamical method, an important deviation around 16$\%$ exists. In addition, it is notable that while the central values of the energy density of the thermodynamical method are similar, in the dynamical one, the full expression leads to slightly higher values. In Table~\ref{tab-Mmin} we compare also the results found by employing the remaining TC$_{i}$ matching processes. In particular, we found that the deviation of the minimum mass due to the method is almost of the same order as the deviation due to the matching process. In other words, one can hardly determine the magnitude of the effect of the method on the value of the minimum mass. It is worth noticing that according to Tables~\ref{tab:1} and~\ref{tab:2}, the increase of the transition density leads to more accurate results between the methods and the matching processes in the case of the minimum mass. However, we found that the effect of the method on the values of $R_{\rm min}$ is more pronounced since the deviation is around $10-14\%$, where the deviation due to the matching process is less than $5\%$. Finally, in the case of the central energy density, the effects of the method are moderate with a deviation around $2\%$ compared to the less than $1\%$ due to the matching process.

It is concluded that the effects of the crust-core interface on the minimum mass  are indistinguishable, are moderate for the  central density, but  concerning the radius are  not negligible. Since the values of the central densities are even lower than those of the value of the saturation density $\rho_s \simeq 2.7\times 10^{14}~{\rm gr~cm^{-3}}$, a final comment is appropriate. In this case, the structure of the core resembles a huge finite nucleus and, thus, makes it an astronomical laboratory to check  properties of low density nuclear matter. In view of the above statement, it will be of interest to study additional effects on the minimum mass configuration, including thermal and rotation effects, and moreover to relate them with known properties of finite nuclei.

\section{Concluding remarks} \label{sec:remarks}
The values of the transition density and pressure are sensitive both to the order of the expansion of the total energy around the asymmetry parameter $I$, and also to the employed method to locate the latter values. Moreover, we found that the lower the value of $L$ the lower the deviation of the results. For higher values of $L$, the deviation becomes appreciable and must be taken into account in order to ensure the accuracy of applications. It is notable that using the full expression the prediction of the transition density, in the case of the dynamical method, satisfies a kind of universal relation with the slope parameter which has been already suggested by Steiner \textit{et al.}~\cite{Steiner-2015}. The corresponding predictions of the thermodynamical method are shifted to slightly higher values. These results confirm, once again, that the dynamical method is more complete and consequently, more accurate compared to the thermodynamical one.
     
The latter results for $n_{\rm t}$ and $P_{\rm t}$ have been applied for the predictions of the bulk neutron star properties, which are related directly to the location of the crust-core transition. According to our findings, the effects are more pronounced on the crustal moment of inertia and the critical frequency related to $r$-mode instabilities. Moreover, the effects can been observed on the estimation of the effective tidal deformability $\tilde{\Lambda}$ (less pronounced on the estimation of the tidal deformability $\lambda$ and tidal Love number $k_2$ for a single neutron star). There are two main remarks for the transition density region that we focused on in our paper ($n_{\mathrm {t}}<0.1\;\mathrm{fm^{-3}}$). First, the parabolic approximations lead to higher values of $\tilde{\Lambda}$ compared to the full expressions. Secondly, the matching process also affects the $\tilde{\Lambda}$, i.e., the TC$_1$ matching process slightly shifts the curves to lower values of $\tilde{\Lambda}$. Also, according to our paper, gravitational wave events of binary neutron star mergers with lower component masses (hence lower chirp mass $\mathcal{M}_c$) might be more suitable as a tool for distinguishing the different methods and matching processes. Furthermore, in minimum mass configuration, while the effects are almost imperceptible for the minimum gravitational mass and the corresponding central energy density, a significant effect is presented in the corresponding radius.

Finally, we conclude that the dynamical method, in the framework of the full expression of the total energy, leads to more accurate predictions for the relevant neutron star properties. In contrast, the most used thermodynamical method in the framework of the parabolic approximation is responsible for the roughest predictions.

{\it Note added.} Recently, a related work, with some similar  predictions and conclusions, has appeared~\cite{Souleiman-2021}.

\section*{Acknowledgments}
The authors would like to thank Prof. K. Kokkotas for his useful comments and insight and  Dr. S. Typel for useful discussions concerning the matching process at the phase transition. We also thank the anonymous reviewer for valuable comments and suggestions.

\section*{Appendix: Matching process} \label{sec:appendix}
We employ a thermodynamic consistency method in order to match the two EoSs (those correspond to the core and the crust). In particular, the method is based on the $P(n)$ and $n$ relation, where we consider the baryon density $n$ as an independent variable. A detailed presentation of the method is given in Ref.~\cite{Fortin-2016}. In the first region, which corresponds to the crust (hereafter denoted with the index 1), the relation $P_{\rm cr}(n)\equiv P_1(n)$ holds. In the second one, which corresponds to the core (hereafter denoted with the index 2), the relation $P_{\rm core}(n)\equiv P_2(n)$ holds. Moreover,  we consider that the matching region (mr) lies between the two densities $n_1$ and $n_2$, where $n_2>n_1$, and the EoS is denoted as $P_{\rm mr}(n)$. For the matching region we employ a linear dependence of $P_{\rm mr}(n) $ on $n$ and by considering the continuity relations, $P_{\rm mr}(n_1)=P_1(n_1)$ and  $P_{\rm mr}(n_2)=P_2(n_2)$, we found for the matching region the relation
\begin{equation}
	P_{\rm mr}(n)=P_1(n_1)+\alpha(n-n_1),
	\label{ap-1}
\end{equation}
where
\[ \alpha=\frac{P_2(n_2)-P_1(n_1)}{n_2-n_1}. \]
In the matching region and considering that ${\cal E}_{\rm mr}(n)=n\mu_{\rm mr}(n)-P_{\rm mr}(n)$ (where  $\mu=d{\cal E}/{dn}$), the chemical potential $\mu_{\rm mr}(n)$ is given by
\begin{equation}
	\mu_{\rm mr}(n)=\mu_1(n_1)+\int_{n_1}^{n}\frac{dP_{\rm mr}(n)}{n},
	\label{ap-2}
\end{equation}
where $\mu_1(n_1)=(P_1+{\cal E}_1)/n_1$ and 
\begin{equation}
	\mu(n_2)=\mu_1(n_1)+\int_{n_1}^{n_2}\frac{dP_{\rm mr}(n)}{n}.
	\label{ap-2}
\end{equation}
However, in general, $\mu_{\rm mr}(n_2)\neq \mu_2(n_2)$ (where $\mu_2=(P_2+{\cal E}_2)/n_2)$. In this case, in order to satisfy the thermodynamically consistent EoS for $n>n_2$, we define the difference $\Delta\mu= \mu_{\rm mr}(n_2)-\mu_2(n_2)$. 

Summarizing, the EoSs of each of the three regions, are specified as follow:
\begin{itemize}
\item[(1)] For the crust ($n<n_1$)  we consider the EoS given in Ref.~\cite{Baym-71b}
    
\item[(2)] For the matching region ($n_1<n<n_2$) the pressure is given by Eq.~\eqref{ap-1} and the energy density by ${\cal E}_{\rm mr}(n)=n\mu_{\rm mr}(n)-P_{\rm mr}(n)$ where
\begin{equation}
	\mu_{\rm mr}(n)=\mu_1(n_1)+ \alpha\ln \left(\frac{n}{n_1}\right),
	\label{ap-3}
\end{equation} 
and also
\begin{eqnarray}
	&&P_{\rm mr}(\mu)=P_1(n_1)+\alpha n_1\left(e^{\frac{\mu-\mu_1}{\alpha}}-1  \right). 
 	\label{ap-4}
\end{eqnarray}
 
\item[(3)] For the core ($n>n_2$) we employ the pressure $P_2(n)$ which is taken from the MDI(80) model while the corresponding energy density ${\cal E}_2(n)$ will be given by ${\cal E}_2(n)={\cal E}(n)+n\Delta\mu$ [where ${\cal E}(n)$ is the energy density taken from the MDI(80) model].

\end{itemize}

Finally, the speed of sound in the matching region will be given by the expression
\begin{equation}
	\frac{v_{\rm s}}{c} \equiv\sqrt{\frac{\partial P}{\partial {\cal E}}}=\sqrt{\frac{\alpha}{\mu_1+\alpha\ln(\frac{n}{n_1})}}.
	\label{ap-5}
\end{equation}
In the present paper, we employ three different kinds of matching. In the first one, called TC$_{1}$, the selected densities $n_1$ and $n_2$ lie symmetrically among the critical density $n_{\rm t}$ within the crust and the core, respectively. In the second one, called TC$_{2}$, the density of the core $n_2$ is identified as the transition density and $n_1$ lies in the crust, while in the third case, called TC$_{3}$, the crust density $n_1$ is identified with $n_{\rm t}$ and $n_2$ lies in the core.


\end{document}